\documentclass[a4paper,11pt]{article}
\usepackage{jcappub} 
\usepackage{amsmath,amssymb}
\usepackage{mathrsfs} 
\usepackage{physics}
\usepackage{url}
\usepackage{appendix}

\arxivnumber{2610.xxxxx} 
\title{\boldmath Probing Helical Primordial Magnetic Fields via Chiral Gravitational Waves in the LISA-TAIJI Network}

\author[a]{Kazuya Furusawa}
\author[a]{Hiroyuki Tashiro}
\affiliation[a]{Graduate School of Science, Nagoya University, Furo-cho, Chikusa-ku, Nagoya, Aichi, 464-8602, Japan}
\emailAdd{furusawa.kazuya.m8@s.mail.nagoya-u.ac.jp}

\abstract{
Primordial magnetic fields (PMFs) are anticipated to be a cosmological candidate for generating a stochastic gravitational wave background (SGWB) through their anisotropic stress. 
In particular, if parity violation is present in the production of PMFs, it can induce a helical component in the PMFs, leading to circular polarization in the isotropic SGWB.
In this study, we phenomenologically consider parity-violating inflationary magnetogenesis and compute the intensity and circular polarization of the SGWB for several assumed power spectra of helical PMFs. 
Using the planned sensitivity of the LISA-TAIJI network, we calculate the signal-to-noise ratio (SNR) and perform a Fisher forecast to assess the feasibility of constraining the parameters of helical PMFs.
We conclude that it is possible to estimate the helical-to-non-helical power ratio~$r_H$ with $\mathrm{SNR}^{V}>2$ if the PMF strength is $\mathcal{B}\gtrsim 10~\mathrm{nG}$ ($50~\mathrm{nG}$) in the case of the delta-function-type (scale-invariant-type) PMF spectrum.
Our results offer a useful criterion that indicates the future observational limit on helical PMFs at the small scale $k_\mathrm{LISA} \approx 10^{12}~\mathrm{Mpc}^{-1}$.
}

\begin{document}
\maketitle
\flushbottom

\section{Introduction}

The first direct detection of Gravitational Waves (GWs) by the ground-based detector LIGO in 2015 marked the dawn of GW astronomy, which opened a new window to explore new physics in the universe~\cite{LIGOScientific:2016aoc}. Due to their extremely weak interaction with matter, GWs can be a critical tool to probe the early universe, which is inaccessible to radio observations~\cite{Maggiore:2007ulw}. 

Most cosmological GW sources, such as inflationary GWs, cosmic strings, and first-order phase transitions, are considered to generate a stochastic GW background (SGWB).
Therefore, searching for SGWB potentially gives us valuable insight into phenomena beyond standard cosmology and particle physics (see e.g. Ref.~\cite{Caprini:2018mtu, Maggiore:2018sht} for reviews). 

One possible source of GWs in the early universe is primordial magnetic fields (PMFs). 
PMFs are weak magnetic fields generated in the early Universe, which are thought to have been produced during cosmic inflation or the electroweak phase transition (see e.g. Ref.~\cite{Durrer:2013pga,Subramanian:2015lua} for reviews).
They are widely regarded as the seeds of the cosmic magnetic fields observed in galaxies and galaxy clusters.
In particular, PMFs are promising candidates for naturally explaining the presence of magnetic fields in cosmic voids inferred from the observations of TeV blazars by the Fermi Large Area Telescope~\cite{Tavecchio:2010mk, Neronov:2010gir}. If PMFs exist in the early universe, PMFs affect spacetime through their anisotropic stress, leading to the generation of SGWB~\cite{Kosowsky:2001xp,Mack:2001gc,Saga:2018ont}. 
Although PMFs on small scales are expected to be damped due to the back-reaction from the cosmic plasma in the evolution of the Universe~\cite{Jedamzik:1996wp, Subramanian:1997gi}, the resulting SGWB remains as a long-lasting imprint of these fields and survives to the present epoch. 
Thus, GW observations can also offer the valuable opportunity to reveal the origin of the cosmic magnetic field in the present universe.

An important property of PMFs is their helicity.  
Helical PMFs naturally arise in several generation scenarios involving parity-violating interactions during inflation~\cite{Caprini:2014mja, Fujita:2019pmi,Sharma:2018kgs}.  
Such fields can undergo an inverse cascade, transferring magnetic energy from small to large scales, which allows helical components to survive longer than non-helical ones~\cite{Christensson:2000sp, Banerjee:2004df}. 
This long-term survival and large-scale coherence make helical PMFs particularly appealing as seed fields for the magnetic fields observed in galaxies, galaxy clusters, and even cosmic voids, potentially contributing to their present-day strength and structure~\cite{Tashiro:2013ita}.
Because magnetic helicity is directly related to parity violation, the detection of its cosmological signatures would provide key evidence for parity-violating processes in the early Universe.

If PMFs possess nonzero helicity, they can act as a parity-violating source and generate circularly polarized gravitational waves~\cite{Caprini:2003vc,Kahniashvili:2005qi}.  
Circular polarization of GWs represents an asymmetry between the left-handed and right-handed polarization modes, and is therefore regarded as a promising observable for probing parity-violating physics in the early Universe.  
Detecting or constraining the degree of circular polarization in the SGWB would thus offer a direct observational test of the helicity of primordial magnetic fields and, more broadly, of parity-violating physics in the early Universe.  
In this context, helical PMFs provide a particularly compelling mechanism linking early-Universe magnetogenesis, parity violation, and the observable circular polarization of the stochastic GW background~\cite{Okano:2020uyr, Sharma:2019jtb, Brandenburg:2021bfx}.

In particular, a joint international observational project, the LISA-TAIJI network, has been proposed in Refs.~\cite{Ruan:2020smc,Wang:2021uih}.  
This project aims to combine the observational data from the two spatially separated detectors during their overlapping operational periods.  
Such a configuration enables the measurement of the circular polarization of an isotropic SGWB, which cannot be accessed by a single planar detector alone~\cite{Chen:2024ikn,Chen:2024fto}.  
Therefore, future space-based GW missions such as LISA and TAIJI hold great promise for directly probing parity-violating signatures imprinted in the early Universe, including helical PMFs.

In this study, we phenomenologically consider parity-violating inflationary magnetogenesis.

We compute the intensity and circular polarization of SGWB for various assumed power spectra of helical PMFs and compare them with the planned sensitivity of the LISA-TAIJI network given in Ref.~\cite{Chen:2024ikn}.
Finally, we discuss the possibility of probing this parity-violating signature associated with helical PMFs by the LISA-TAIJI network. 
Note that the chiral GWs in our scenario behave as the passive tensor modes considered in Refs.~\cite{Shaw:2009nf,Planck:2015zrl}, rather than those generated by helical MHD turbulence in Ref.~\cite{RoperPol:2021xnd, Kahniashvili:2020jgm}.

The structure of this paper is as follows. In Section~\ref{sec:generation}, we explain how helical PMFs source a parity-violating anisotropic stress. 
In Section~\ref{sec:polarized-sgwb}, we derive the resulting circularly polarized SGWB spectra. 
In Section~\ref{sec:lisa-taiji}, we introduce the formulation of the LISA-TAIJI network. 
In Section~\ref{sec:detectability}, we evaluate the detectability of SGWB from helical PMFs. Finally, in Section~\ref{sec:summary}, we summarize this paper.
In our calculation, we adopt the flat~$\Lambda\mathrm{CDM}$~model with $h=0.7$, $\Omega_\mathrm{m}=0.3$ and $\Omega_\Lambda =0.7$
together with the photon fraction of the radiation density, $\rho_\gamma/\rho_r \approx 0.6$~(assuming the standard value $N_{\rm eff} =3.046$).

\section{Generation of helical PMFs and their anisotropic stress}
\label{sec:generation}

In this section, we introduce the statistical properties of helical primordial
magnetic fields (PMFs) and derive the resulting anisotropic stress that acts
as a source for gravitational waves. We first define the symmetric and helical
components of the PMF power spectrum in Sec.~\ref{sec:helical-pmf}, and then
project the magnetic energy-momentum tensor onto its parity-even and
parity-odd components in Sec.~\ref{sec:aniso-stress}. Finally, in Sec.~\ref{sec:pheno-models}, we specify two
phenomenological PMF models that we use throughout the rest of this paper. Throughout this section, we adopt units in which the speed of light is unity, $c=1$, while we write $c$ explicitly in the other sections.
We also normalize the scale factor to unity today, $a_0=1$, so that the comoving quantities introduced below~(e.g., ${\tilde B}_i$)
coincide with their present-day physical values.

\subsection{Helical primordial magnetic fields}
\label{sec:helical-pmf}

We assume that PMFs are statistically homogeneous and isotropic Gaussian
random fields with non-vanishing magnetic helicity. In the highly conductive
primordial plasma, the physical magnetic field evolves adiabatically as
$B_i(\vb*{x},\eta)\propto a^{-2}(\eta)$. It is therefore convenient to
introduce the comoving magnetic field,
\begin{equation}
  \tilde B_i(\vb*{x}) \equiv a^2(\eta)\, B_i(\vb*{x},\eta),
  \label{eq:comoving-B}
\end{equation}
which remains constant in time in the ideal magnetohydrodynamic (MHD) limit.
The Fourier transform of the comoving magnetic field is defined as
\begin{equation}
  \tilde B_i(\vb*{x}) = \int \frac{d^3k}{(2\pi)^3}\, \tilde B_i(\vb*{k})\, e^{i\vb*{k}\cdot\vb*{x}}.
  \label{eq:B-fourier}
\end{equation}

For statistically homogeneous and isotropic helical PMFs, the power spectrum
of the comoving magnetic field is written as
\begin{equation}
  \left\langle \tilde B_i(\vb*{k})\, \tilde B_j^*(\vb*{k}') \right\rangle
  = \frac{(2\pi)^3}{2} \delta_\mathrm{D}^{(3)}(\vb*{k}-\vb*{k}')
  \left[ P_{ij}(\hat{\vb*{k}})\, S(k) + i\,\epsilon_{ijl}\hat k_l\, A(k) \right],
  \label{eq:B-power-spectrum}
\end{equation}
where $\delta_\mathrm{D}$ is the delta-function and 
\begin{equation}
  P_{ij}(\hat{\vb*{k}}) = \delta_{ij} - \hat k_i \hat k_j
  \label{eq:transverse-projector}
\end{equation}
is the transverse projection tensor. The symmetric component $S(k)$
represents the parity-even magnetic power spectrum, while the antisymmetric
component $A(k)$ characterizes the magnetic helicity and encodes parity
violation. The realizability condition requires
\begin{equation}
  |A(k)| \le S(k),
  \label{eq:realizability}
\end{equation}
which follows from the Cauchy-Schwarz inequality~\cite{Durrer:2003ja}. In what follows, we keep the spectral shapes of $S(k)$ and $A(k)$ arbitrary and
develop a general formalism for gravitational-wave production from helical
PMFs; specific phenomenological choices are introduced in
Sec.~\ref{sec:pheno-models}.

\subsection{Anisotropic stress from PMFs}
\label{sec:aniso-stress}

In cosmological linear perturbation theory, PMFs generate gravitational waves
through the transverse-traceless (TT) part of their anisotropic stress. Since
the magnetic energy-momentum tensor is quadratic in the magnetic field and
$B_i\propto a^{-2}$, the anisotropic stress scales as
$\Pi^{(B)}_{ij}\propto a^{-4}$. Following Refs.~\cite{Shaw:2009nf,Saga:2018ont}, it is convenient to normalize the anisotropic
stress tensor by the photon pressure $p_\gamma=\rho_\gamma/3$ and define the
dimensionless comoving anisotropic stress tensor $\tilde\Pi_{ij}$ as
\begin{equation}
  \Pi^{(B)}_{ij}(\vb*{k},\eta) = p_\gamma(\eta)\, \tilde{\Pi}_{ij}(\vb*{k}).
  \label{eq:aniso-stress-def}
\end{equation}
Since $p_\gamma\propto\rho_\gamma\propto a^{-4}$, the quantity $\tilde\Pi_{ij}$ is time
independent in the ideal MHD limit. It is obtained by projecting the magnetic
energy-momentum tensor onto its TT part,
\begin{equation}
  \tilde\Pi_{ij}(\vb*{k}) = \Lambda_{ij,lm}(\hat{\vb*{k}})\, T^{(\tilde B)}_{lm}(\vb*{k}),
  \label{eq:Pi-projection}
\end{equation}
where
\begin{equation}
  T^{(\tilde B)}_{ij}(\vb*{k}) = \frac{3}{4\pi\rho_{\gamma,0}}
  \int \frac{d^3p}{(2\pi)^3}
  \left[ \tilde B_i(\vb*{p})\tilde B_j(\vb*{k}-\vb*{p})
  - \frac{1}{2}\delta_{ij}\tilde B_l(\vb*{p})\tilde B_l(\vb*{k}-\vb*{p}) \right]
  \label{eq:T-Btilde}
\end{equation}
is the magnetic energy-momentum tensor normalized by the present photon
energy density $\rho_{\gamma,0}$, and
$\Lambda_{ij,lm}(\hat{\vb*{k}}) \equiv P_{il}P_{jm} - \tfrac{1}{2}P_{ij}P_{lm}$
is the TT projection operator.

Because $T^{(\tilde B)}_{ij}$ is quadratic in $\tilde B_i$, its two-point
function $\langle \tilde\Pi_{ij}\tilde\Pi_{lm}^*\rangle$ involves the
four-point function of the Gaussian field $\tilde B_i$. Applying Wick's
theorem, this reduces to products of the two-point function
\eqref{eq:B-power-spectrum} evaluated at $\vb*{p}$ and $\vb*{k}-\vb*{p}$,
integrated over the loop momentum $\vb*{p}$.

Following
Refs.~\cite{Caprini:2003vc,Saga:2018ont}, the resulting two-point correlation
function of $\tilde\Pi_{ij}$ can be decomposed into parity-even and
parity-odd components as
\begin{equation}
  \left\langle \tilde\Pi_{ij}(\vb*{k})\, \tilde\Pi_{lm}^*(\vb*{k}') \right\rangle
  = \frac{(2\pi)^3}{4} \delta_\mathrm{D}^{(3)}(\vb*{k}-\vb*{k}')
  \left[ \mathcal{M}_{ijlm}(\hat{\vb*{k}})\, f(k) + i \mathcal{A}_{ijlm}(\hat{\vb*{k}})\, g(k) \right],
  \label{eq:Pi-correlator}
\end{equation}
where
\begin{align}
  \mathcal{M}_{ijlm}(\hat{\vb*{k}}) &= P_{il}P_{jm} + P_{im}P_{jl} - P_{ij}P_{lm}, \\
  \mathcal{A}_{ijlm}(\hat{\vb*{k}}) &= \frac{\hat k_q}{2}
  \left( P_{jm}\epsilon_{ilq} + P_{il}\epsilon_{jmq}
  + P_{im}\epsilon_{jlq} + P_{jl}\epsilon_{imq} \right).
  \label{eq:M-A-tensors}
\end{align}
The functions $f(k)$ and $g(k)$ represent the parity-even and parity-odd
components of the anisotropic stress spectrum, respectively. Projecting the
magnetic-field correlator \eqref{eq:B-power-spectrum} via the Wick
contraction described above, they are given by
\begin{align}
  f(k) &= \frac{3^2}{4(4\pi)^2\rho_{\gamma,0}^2} \int \frac{d^3p}{(2\pi)^3}
  \Big[ (1+\gamma^2)(1+\beta^2)\, S(p)\,S(|\vb*{k}-\vb*{p}|)
  + 4\gamma\beta\, A(p)\,A(|\vb*{k}-\vb*{p}|) \Big], \label{eq:f-general} \\
  g(k) & = \frac{3^2}{2(4\pi)^2\rho_{\gamma,0}^2} \int \frac{d^3p}{(2\pi)^3}(1+\gamma^2)\,\beta\, S(p)\,A(|\vb*{k}-\vb*{p}|)\gamma(1+\beta^2)\, A(p)\,S(|\vb*{k}-\vb*{p}|), \label{eq:g-general}
\end{align}
with $\gamma=\hat{\vb*{k}}\cdot\hat{\vb*{p}}$ and
$\beta=\hat{\vb*{k}}\cdot\widehat{\vb*{k}-\vb*{p}}$, and $k\equiv|\vb*{k}|$,
$p\equiv|\vb*{p}|$.

The parity-even function $f(k)$ sources gravitational waves equally in the
two helicity states, whereas the parity-odd function $g(k)$ induces an
asymmetry between right- and left-handed gravitational waves. A non-vanishing
helical component $A(k)$ is therefore both necessary and sufficient for
$g(k)\ne0$, and hence for parity violation in the resulting stochastic
gravitational-wave background (SGWB); this is made explicit once we project
Eq.~\eqref{eq:Pi-correlator} onto the circular polarization basis in
Sec.~\ref{sec:polarized-sgwb}.

\subsection{Phenomenological PMF models}
\label{sec:pheno-models}

To evaluate the general formulas of Sec.~\ref{sec:aniso-stress} concretely, we
consider two representative phenomenological PMF spectra, following
Ref.~\cite{Saga:2018ont}: a delta-function-type spectrum and a scale-invariant
spectrum. The resulting SGWB spectra for these two models are presented in
Sec.~\ref{sec:pheno-spectra}; their derivation is given in Appendix~\ref{app:fg-derivation}.

\subsubsection{delta-function-type PMFs}
\label{sec:delta-type}

The delta-function-type PMF spectrum is defined as
\begin{align}
  S(k) &= \frac{2\pi^2}{k^3}\mathcal{B}^2 \delta_\mathrm{D}\!\left[\left(\ln\frac{k}{k_\mathrm{p}}\right)\right], \\
  A(k) &= \frac{2\pi^2}{k^3} r_H\mathcal{B}^2\delta_\mathrm{D}\!\left[\left(\ln\frac{k}{k_\mathrm{p}}\right)\right],
  \label{eq:delta-type}
\end{align}
where $\mathcal{B}$ denotes the comoving PMF amplitude, $k_\mathrm{p}$ is the
characteristic PMF scale, and $r_H$ parameterizes the fractional magnetic
helicity introduced in Ref.~\cite{Yura:2025mus}. The realizability condition~\eqref{eq:realizability} requires
$|r_H|\le 1$. This spectrum corresponds to PMFs concentrated around the
characteristic scale $k=k_\mathrm{p}$.

\subsubsection{Scale-invariant PMFs}
\label{sec:scale-invariant-type}

Motivated by inflationary magnetogenesis scenarios~\cite{Fujita:2019pmi}, we
also consider a scale-invariant PMF spectrum,
\begin{align}
  S(k) &= \frac{2\pi^2}{k^3}\mathcal{B}^2 \left[\ln\!\left(\frac{k_{\max}}{k_{\min}}\right)\right]^{-1}, \\
  A(k) &= \frac{2\pi^2}{k^3}\mathcal{B}^2 r_H \left[\ln\!\left(\frac{k_{\max}}{k_{\min}}\right)\right]^{-1},
  \label{eq:scale-inv-type}
\end{align}
for $k_{\min}\le k\le k_{\max}$, and vanishing outside this range. The
normalization factor $[\ln(k_{\max}/k_{\min})]^{-1}$ is introduced so that
$\mathcal{B}^2$ corresponds to the variance of the comoving magnetic field. To
avoid a logarithmic divergence, infrared and ultraviolet cut-offs are imposed.
Following the validity range of the transfer-function
approximation~\eqref{eq:Th-eq-approx} (see Sec.~\ref{sec:transfer-function}),
we adopt $k_{\min}=\eta_\nu^{-1}$ with he conformal time at neutrino
decoupling, $\eta_\nu\approx7.6\times10^{-4}~\mathrm{Mpc}$.
On the other hand, for the ultraviolet cut-off we choose
$k_{\max}=10^8\,\eta_B^{-1}$, 
where $\eta_B$ denotes the conformal time at PMF generation.
In this paper, we assume an inflationary magnetic-genesis scenario in which PMFs are generated at the end of inflation $\eta = \eta_B$, corresponding to the beginning of the radiation-dominated era.

\section{Circularly polarized stochastic gravitational wave background}
\label{sec:polarized-sgwb}

In this section, 
we show how the anisotropic stress due to the helical PMFs
sources circularly polarized gravitational waves, and compute the resulting energy density spectra. 
We first project the tensor perturbation and its
source onto the circular polarization basis and derive the Stokes parameters
$I(k,\eta)$ and $V(k,\eta)$ in Sec.~\ref{sec:chiral-gw}. We then specify the
transfer function relevant for present-day observations in Sec.~\ref{sec:transfer-function} 
and define the observable energy density
spectra $\Omega_\mathrm{gw}^I(k)$ and $\Omega_\mathrm{gw}^V(k)$ in Sec.~\ref{sec:energy-density}.
Finally, in  Sec.~\ref{sec:pheno-spectra}, we evaluate these spectra for the two phenomenological PMF models
introduced in Sec.~\ref{sec:pheno-models}.

\subsection{Chiral gravitational waves from anisotropic stress}
\label{sec:chiral-gw}

In cosmological perturbation theory, gravitational waves are described by
tensor perturbations $h_{ij}$ of the spatial part of the
Friedmann--Lema\^itre--Robertson--Walker metric,
\begin{equation}
  ds^2 = a^2(\eta) \left[ -d\eta^2 + (\delta_{ij}+h_{ij})\,dx^i dx^j \right],
  \label{eq:frw-metric}
\end{equation}
where the tensor perturbation satisfies the transverse-traceless conditions
$\partial_i h_{ij}=0$, $h_{ii}=0$. In Fourier space, the evolution equation
for tensor perturbations is
\begin{equation}
  h_{ij}''(\vb*{k},\eta) + 2\mathcal{H}\, h_{ij}'(\vb*{k},\eta) + k^2 h_{ij}(\vb*{k},\eta)
  = 16\pi G\, a^2(\eta)\, \rho_\gamma(\eta)\, \tilde\Pi_{ij}(\vb*{k}),
  \label{eq:tensor-eom}
\end{equation}
where a prime denotes the derivative with respect to conformal time and
$\mathcal{H}\equiv aH =  a'/a$. During the radiation-dominated era, writing
$\rho_\gamma(\eta) = R_\gamma\, \rho_r(\eta)$ with $R_\gamma\equiv\rho_\gamma/\rho_r$~(the photon fraction of the total radiation energy density), the Friedmann
equation allows us to rewrite Eq.~\eqref{eq:tensor-eom} as
\begin{equation}
  h_{ij}''(\vb*{k},\eta) + 2\mathcal{H}\, h_{ij}'(\vb*{k},\eta) + k^2 h_{ij}(\vb*{k},\eta)
  = 2 R_\gamma \mathcal{H}^2\, \tilde\Pi_{ij}(\vb*{k}).
  \label{eq:tensor-eom-rad}
\end{equation}

To characterize the polarization state of the gravitational waves, we
decompose the tensor perturbation into the circular polarization basis,
\begin{equation}
  h_{ij}(\vb*{k},\eta) = \sum_{\lambda=R,L} e^\lambda_{ij}(\hat{\vb*{k}})\, h_\lambda(k,\eta),
  \label{eq:circular-decomp}
\end{equation}
where the circular polarization tensors satisfy
\begin{equation}
  i\,\epsilon_{ilm}\hat k_l\, e^{R,L}_{mj} = \pm\, e^{R,L}_{ij},
  \label{eq:circular-tensor-def}
\end{equation}
with the plus and minus signs corresponding to the right- and left-handed
polarization states, respectively, and normalized such that
$e^{\lambda*}_{ij}(\hat{\vb*{k}})\, e^{\lambda'}_{ij}(\hat{\vb*{k}}) = 2\,\delta_{\lambda\lambda'}$.
Similarly, the anisotropic stress tensor is projected onto the helicity basis
as
\begin{equation}
  \tilde\Pi_\lambda(\vb*{k}) = e^{\lambda*}_{ij}(\hat{\vb*{k}})\, \tilde\Pi_{ij}(\vb*{k}).
  \label{eq:Pi-helicity-projection}
\end{equation}
Accordingly, the projection of Eq.~\eqref{eq:tensor-eom-rad} onto the circular polarization basis
provide us the following evolution,
\begin{equation}
  h_\lambda''(k,\eta) + 2\mathcal{H}\, h_\lambda'(k,\eta) + k^2 h_\lambda(k,\eta)
  = 2 R_\gamma \mathcal{H}^2\, \tilde\Pi_\lambda(k).
  \label{eq:h-lambda-eom}
\end{equation}
The formal solution can be written using the Green-function method as
\begin{equation}
  h_\lambda(k,\eta) = 2 R_\gamma \int^\eta d\eta'\, G_k(\eta,\eta')\, \mathcal{H}^2(\eta')\, \tilde\Pi_\lambda(k),
  \label{eq:h-lambda-solution}
\end{equation}
where $G_k(\eta,\eta')$ is the Green function associated with
Eq.~\eqref{eq:h-lambda-eom}.

To project the anisotropic-stress correlator \eqref{eq:Pi-correlator} onto
this basis, we use the contraction identities
\begin{equation}
  e^{\lambda*}_{ij}(\hat{\vb*{k}})\, e^{\lambda'}_{lm}(\hat{\vb*{k}})\, \mathcal{M}_{ijlm}(\hat{\vb*{k}}) = 4\,\delta_{\lambda\lambda'},
  \qquad
  e^{\lambda*}_{ij}(\hat{\vb*{k}})\, e^{\lambda'}_{lm}(\hat{\vb*{k}})\, \mathcal{A}_{ijlm}(\hat{\vb*{k}}) = 4\lambda\,\delta_{\lambda\lambda'},
  \label{eq:MA-orthogonality}
\end{equation}
which follow directly from the defining property~\eqref{eq:circular-tensor-def}
of the circular polarization tensors. Contracting
Eq.~\eqref{eq:Pi-correlator} with $e^{\lambda*}_{ij}e^{\lambda'}_{lm}$ and
using Eq.~\eqref{eq:MA-orthogonality} then gives
\begin{equation}
 \left\langle \tilde\Pi_\lambda(\vb*{k})\, \tilde\Pi_{\lambda'}^*(\vb*{k}') \right\rangle
  = \frac{(2\pi)^3}{4} \delta_\mathrm{D}^{(3)}(\vb*{k}-\vb*{k}')\, \delta_{\lambda\lambda'}
  \left[ f(k) + \lambda\, g(k) \right],
  \label{eq:Pi-lambda-correlator}
\end{equation}
where $\lambda=+1$ and $-1$ correspond to the right- and left-handed
polarization states, respectively.

The statistical properties of the SGWB are characterized by the two-point
correlation function
\begin{equation}
  \left\langle h_\lambda(\vb*{k},\eta)\, h_{\lambda'}^*(\vb*{k}',\eta) \right\rangle
  = (2\pi)^3 \delta_\mathrm{D}^{(3)}(\vb*{k}-\vb*{k}')\, \delta_{\lambda\lambda'}\,\mathcal{P}_\lambda(k,\eta),
  \label{eq:h-power-spectrum-def}
\end{equation}
where $\mathcal{P}_R(k,\eta)$ and $\mathcal{P}_L(k,\eta)$ denote the power
spectra of the right- and left-handed gravitational waves, respectively.
Since $\tilde\Pi_{ij}$ is time-independent in the ideal MHD limit, the
gravitational-wave power spectra can be expressed as
\begin{equation}
  \mathcal{P}_\lambda(k,\eta) = \frac{1}{4}\,|T_h(k,\eta)|^2\, R_\gamma^2 \left[ f(k) + \lambda\, g(k) \right],
  \label{eq:P-lambda}
\end{equation}
where $T_h(k,\eta)$ denotes the transfer function associated with the
Green-function integration in Eq.~\eqref{eq:h-lambda-solution},
\begin{equation}
  T_h(k,\eta) = 2 \int^\eta d\eta'\, G_k(\eta,\eta')\, \mathcal{H}^2(\eta').
  \label{eq:Th-def}
\end{equation}
Hence the parity-even source spectrum $f(k)$ contributes equally to the two
helicity states, whereas the parity-odd source spectrum $g(k)$ induces an
asymmetry between the right- and left-handed modes:
\begin{align}
  \mathcal{P}_R(k,\eta) &=  \frac{1}{4}\,|T_h(k,\eta)|^2\, R_\gamma^2 \left[ f(k) + g(k) \right], \label{eq:PR} \\
  \mathcal{P}_L(k,\eta) &= \frac{1}{4}\,|T_h(k,\eta)|^2\, R_\gamma^2 \left[ f(k) - g(k) \right]. \label{eq:PL}
\end{align}

In analogy with electromagnetic waves, the intensity and circular
polarization of the SGWB are characterized by the Stokes parameters
\begin{align}
  \frac{2\pi}{k^2}I(k,\eta) &= \mathcal{P}_R(k,\eta) + \mathcal{P}_L(k,\eta)
  = \frac{1}{2}\,|T_h(k,\eta)|^2\, R_\gamma^2\, f(k), \label{eq:I-stokes}\\
  \frac{2\pi}{k^2}V(k,\eta) &= \mathcal{P}_R(k,\eta) - \mathcal{P}_L(k,\eta)
  = \frac{1}{2}\,|T_h(k,\eta)|^2\, R_\gamma^2\, g(k).\label{eq:V-stokes}
\end{align}
The intensity parameter $I(k,\eta)$ characterizes the total power spectrum of
the SGWB, while the circular polarization parameter $V(k,\eta)$ characterizes
the asymmetry of the power spectrum between right- and left-handed gravitational waves. A
non-vanishing $V(k,\eta)$ therefore indicates parity violation in the
gravitational-wave background, and by Eq.~\eqref{eq:V-stokes} directly traces
the helical PMF source $g(k)$.
Here we multiply $I(k,\eta)$ and $V(k,\eta)$ by the factor $2\pi/k^2$ to keep consistency with the conventional definition of the Stokes parameters $I(f)$ and $V(f)$ in the GW detection statistic described in Sec.~\ref{sec:lisa-taiji} and Refs.~\cite{Chen:2024ikn,Chen:2024fto}.

\subsection{Transfer function of PMF-induced gravitational waves}
\label{sec:transfer-function}

To evaluate the present-day SGWB spectra, we specify the transfer function $T_h(k,\eta_0)$ that describes the evolution of tensor perturbations after production. As discussed in~\cite{Saga:2018ont},
Gravitational waves are continuously sourced by the magnetic
anisotropic stress first at super-horizon scales during the radiation-dominated era, with sourcing most efficient around horizon entry, $k\eta\sim1$, where the Green-function integration in Eq.~\eqref{eq:Th-def} efficiently amplifies the tensor perturbations. 
After horizon entry, the source contribution becomes subdominant, and the gravitational waves subsequently evolve approximately as freely propagating tensor modes.

For the modes relevant to space-based gravitational-wave detectors such as
LISA, the tensor perturbations are already well inside the horizon during the
matter-dominated era. Their late-time evolution can therefore be approximated
by the standard adiabatic scaling of a freely propagating tensor mode deep
inside the horizon,
\begin{equation}
  h_\lambda(k,\eta) \simeq \frac{a_{\rm eq}}{a(\eta)}\, h_\lambda(k,\eta_{\rm eq}),
  \qquad \eta \ge \eta_{\rm eq},
  \label{eq:adiabatic-scaling}
\end{equation}
where $a_{\rm eq}$ is the scale factor at matter--radiation equality. Applying
this scaling to the transfer function~\eqref{eq:Th-def} evaluated today,
\begin{equation}
  T_h(k,\eta_0) \simeq a_{\rm eq}\, T_h(k,\eta_{\rm eq}),
  \label{eq:Th-scaling}
\end{equation}
where $T_h(k,\eta_{\rm eq})$ denotes the tensor transfer function evaluated
at matter--radiation equality, including the sourcing effect from PMF
anisotropic stress during the radiation-dominated era. Following
Ref.~\cite{Saga:2018ont}, we adopt the approximate form
\begin{equation}
  |T_h(k,\eta_{\rm eq})|^2 \simeq
  \frac{ [{\rm Ci}(k\eta_{\rm eq})-{\rm Ci}(k\eta_B)]^2 + [{\rm Si}(k\eta_{\rm eq})-{\rm Si}(k\eta_B)]^2 }{(k\eta_{\rm eq})^2},
  \label{eq:Th-eq-approx}
\end{equation}
where ${\rm Ci}(x)$
and ${\rm Si}(x)$ are the cosine- and sine-integral functions, respectively.
This expression is obtained by averaging over the oscillatory behavior of
subhorizon tensor perturbations after horizon entry, and is valid for
$k\gtrsim\eta_\nu^{-1}$. Since the frequency range relevant for space-based
gravitational-wave detectors such as LISA satisfies this condition,
Eq.~\eqref{eq:Th-eq-approx} provides an accurate description of the
present-day SGWB spectra considered in this work.

\subsection{Energy density spectra of the polarized SGWB}
\label{sec:energy-density}

The observable quantity relevant for gravitational-wave experiments is the
energy density spectrum of the SGWB. The energy density of gravitational
waves is given by
\begin{equation}
  \rho_\mathrm{gw} = \frac{1}{32\pi G a^2} \left\langle h_{ij}'(\vb*{x},\eta)\, h_{ij}'(\vb*{x},\eta) \right\rangle,
  \label{eq:rho-GW-def}
\end{equation}
where a prime denotes the derivative with respect to conformal time.
According to the helicity decomposition of Sec.~\ref{sec:chiral-gw}, the
energy density spectrum per logarithmic interval of wavenumber is
\begin{equation}
  \frac{d\rho_\mathrm{gw}}{d\ln k} =
  \frac{k^3}{2\pi^2}\,
  \frac{\langle h_R'(k,\eta)h_R'^*(k,\eta)\rangle + \langle h_L'(k,\eta)h_L'^*(k,\eta)\rangle }{16\pi G a^2}.
  \label{eq:drho-dlnk}
\end{equation}
For gravitational waves observed today, the relevant modes are well inside
the horizon and oscillate approximately as $h_\lambda(k,\eta)\propto e^{ik\eta}$,
so that $\langle h_\lambda' h_\lambda'^*\rangle \simeq k^2 \langle h_\lambda h_\lambda^*\rangle$.
Substituting Eq.~\eqref{eq:h-power-spectrum-def} into
Eq.~\eqref{eq:drho-dlnk} and using Eq.~\eqref{eq:I-stokes}, we obtain
\begin{equation}
  \frac{d\rho_\mathrm{gw}}{d\ln k} = \frac{k^5}{2\pi^2}\, \frac{\langle h_R(k,\eta)h_R^*(k,\eta)\rangle + \langle h_L(k,\eta)h_L^*(k,\eta)\rangle}{16\pi G a^2}= \frac{k^3}{16\pi^2 G a^2}I(k,\eta).
\end{equation}
The corresponding dimensionless energy density spectrum is defined by
$\Omega_\mathrm{gw}^I(k,\eta) \equiv \rho_c^{-1}\, d\rho_\mathrm{gw}/d\ln k$, where
$\rho_c = 3H^2/(8\pi G)$ is the critical energy density. Combining
the above equations, and using Eq.~\eqref{eq:I-stokes},
\begin{equation}
  \Omega_\mathrm{gw}^I(k,\eta) =\frac{1}{12}\Big(\frac{k}{aH}\Big)^2\frac{k^3}{2\pi^2}|T_h(k,\eta)|^2\, R_\gamma^2\, f(k)= \frac{k^3}{6\pi a^2H^2}I(k,\eta)
  \label{eq:Omega-I}
\end{equation}
In the same way, defining the circular-polarization energy density spectrum
as $\Omega_\mathrm{gw}^V(k,\eta) \equiv k^3\, V(k,\eta)/(16\pi^2 a^2H^2)$ and
using Eq.~\eqref{eq:V-stokes},
\begin{equation}
  \Omega_\mathrm{gw}^V(k,\eta) = \frac{1}{12}\Big(\frac{k}{aH}\Big)^2\frac{k^3}{2\pi^2}|T_h(k,\eta)|^2\, R_\gamma^2\, g(k)= \frac{k^3}{6\pi a^2H^2}V(k,\eta)
  \label{eq:Omega-V}
\end{equation}
Therefore, the parity-even source spectrum $f(k)$ determines the intensity
spectrum of the SGWB, whereas the parity-odd source spectrum $g(k)$ generates
the circular-polarization spectrum. A non-vanishing $\Omega_\mathrm{gw}^V$ thus directly
probes the parity violation associated with helical PMFs.

Finally, the present-day observed frequency is related to the comoving
wavenumber through $f = k/(2\pi a_0)$. In the following, we express the SGWB spectra as functions of the observed frequency $\Omega_\mathrm{gw}^{I}(f,\eta_0)\equiv \rho_c^{-1}\, d\rho_\mathrm{gw}/d\ln f$. Since $a_0=1$ and $\dd\ln k = \dd \ln f$, the present SGWB energy density is calculated as
\begin{equation}
  \Omega_\mathrm{gw}^I(f,\eta_0) = \Omega_\mathrm{gw}^{I}(k,\eta_0)\Big|_{k=2\pi f} =  \frac{k^3}{6\pi H_0^2}I(k,\eta_0)\Big|_{k=2\pi f} = \frac{4\pi^2}{3H_0^2}f^3I(f,\eta_0).
\end{equation}
Similarly, the present circular-polarization energy density spectrum $\Omega_\mathrm{gw}^{V}(f,\eta_0)$ is written as
\begin{equation}
  \Omega_\mathrm{gw}^V(f,\eta_0) = \Omega_\mathrm{gw}^{V}(k,\eta_0)\Big|_{k=2\pi f} =  \frac{k^3}{6\pi H_0^2}V(k,\eta_0)\Big|_{k=2\pi f} = \frac{4\pi^2}{3H_0^2}f^3V(f,\eta_0).\\
\end{equation}
By defining $\Omega_\mathrm{gw}^{I, V}(f)\equiv\Omega_\mathrm{gw}^{I,V}(f,\eta_0)$,  $I(f)\equiv I(f,\eta_0)$ and $V(f)\equiv V(f,\eta_0)$, we can obtain the consistent expression of the Stokes parameters $I(f)$ and $V(f)$, and the SGWB energy spectrum $\Omega_\mathrm{gw}^{I,V}(f)$ in the GW detection statistic in Sec.~\ref{sec:lisa-taiji}.

\subsection{Spectra for the phenomenological PMF models}
\label{sec:pheno-spectra}

We now evaluate the general formulas of Secs.~\ref{sec:chiral-gw}--\ref{sec:energy-density}
for the two phenomenological PMF models defined in Sec.~\ref{sec:pheno-models}.
For the delta-function-type spectrum~\eqref{eq:delta-type}, the integrals
\eqref{eq:f-general}--\eqref{eq:g-general} can be evaluated analytically
(Appendix~\ref{app:fg-derivation}), yielding $f(k)\propto \mathcal{B}^4$ and
$g(k)\propto r_H\,\mathcal{B}^4$; the full expressions are given in
Eq.~\eqref{eq:fg-delta} of Appendix~\ref{app:fg-derivation}. For the
scale-invariant spectrum~\eqref{eq:scale-inv-type}, the same integrals are
 numerically evaluated following the method of
Appendix~\ref{app:fg-derivation}, and exhibit the same overall scaling,
$f(k)\propto \mathcal{B}^4$ and $g(k)\propto r_H\,\mathcal{B}^4$.

This scaling is the central result of this section: because $f(k)$ is
quadratic in both $S(p)$ and $A(p)$ while $g(k)$ is linear in $A(p)$, the
SGWB intensity $\Omega_\mathrm{gw}^I$ is mainly controlled by the PMF amplitude
$\mathcal{B}$ and depends only weakly on the helicity fraction $r_H$, whereas
the circular polarization $\Omega_\mathrm{gw}^V$ depends linearly on $r_H$. The circular
polarization of the SGWB therefore provides a direct and comparatively clean
observational probe of the primordial magnetic helicity.

\begin{figure}[t]
  \centering
\includegraphics[width=.47\textwidth]{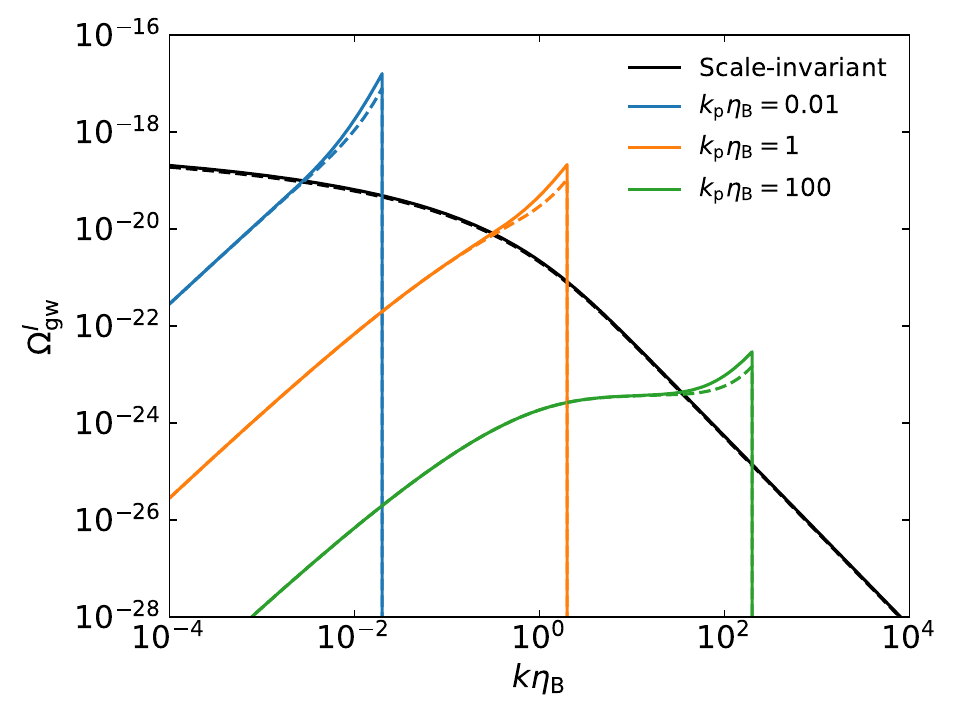}
\qquad
\includegraphics[width=.47\textwidth]{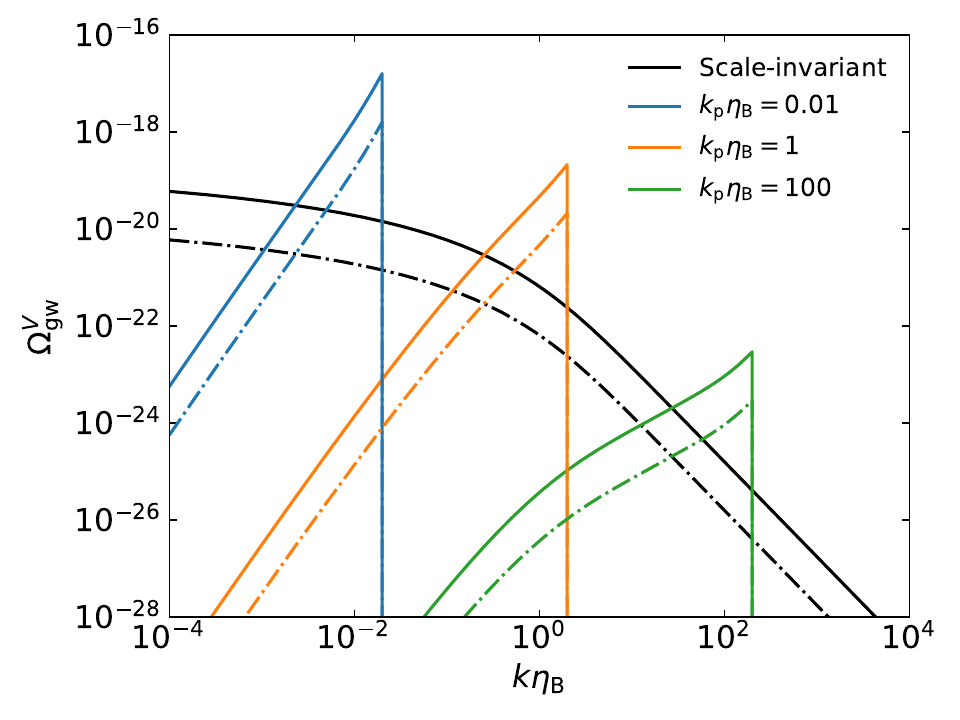}
  \caption{Present-day intensity spectrum $\Omega_\mathrm{gw}^I$ (left) and circular
  polarization spectrum $\Omega_\mathrm{gw}^V$ (right) generated by helical PMFs for the
  phenomenological models of Sec.~\ref{sec:pheno-models}. The comoving
  magnetic-field amplitude is fixed to $\mathcal{B}=1\,{\rm nG}$. For the
  delta-function-type PMFs, three representative peak scales,
  $k_\mathrm{p}\eta_B=0.01,\,1,\,100$, are shown. In the left panel, the solid and
  dashed curves correspond to $r_H=1$ and $r_H=0$, respectively. In the right
  panel, the solid and dash-dotted curves correspond to $r_H=1$ and $r_H=0.1$,
  respectively.}
  \label{fig:omega-spectra}
\end{figure}

Figure~\ref{fig:omega-spectra} shows the resulting present-day spectra
$\Omega_\mathrm{gw}^I$ and $\Omega_\mathrm{gw}^V$ for both models, with the comoving magnetic-field
amplitude fixed to $\mathcal{B}=1\,{\rm nG}$. For the delta-function-type
PMFs, we show three representative peak scales, $k_\mathrm{p}\eta_B=0.01,\,1,\,100$.
The spectra strongly depend on the transfer function of
Sec.~\ref{sec:transfer-function}: PMFs generated on smaller scales source
gravitational waves that enter the horizon earlier, experience a longer
period of subhorizon redshifting, and consequently yield smaller present-day
amplitudes. Compared with the delta-function-type spectrum, the
scale-invariant PMFs generate broader SGWB spectra because magnetic fields
exist over a wide range of scales. In both cases, the left panel confirms
that the dependence of $\Omega_\mathrm{gw}^I$ on $r_H$ is weak, while the right panel
confirms the approximately linear dependence of $\Omega_\mathrm{gw}^V$ on $r_H$,
consistent with the scaling derived above.

\section{LISA--TAIJI network}
\label{sec:lisa-taiji}

In this section, we describe the formalism used to evaluate the
detectability of the polarized SGWB derived in Sec.~\ref{sec:polarized-sgwb}
with the LISA--TAIJI network, following Refs.~\cite{Chen:2024ikn,Chen:2024fto}.
We first introduce the detector response and the cross-correlation estimator
(Sec.~\ref{sec:detector-response}), then derive the signal-to-noise ratio
(SNR) and the power-law integrated (PLI) sensitivity curve
(Sec.~\ref{sec:snr-pli}), and finally introduce the Fisher-matrix formalism
used to forecast parameter constraints (Sec.~\ref{sec:fisher}).

\subsection{Detector response and cross-correlation}
\label{sec:detector-response}

LISA and TAIJI are space-based gravitational-wave interferometers, each
consisting of three spacecraft arranged in a nearly equilateral triangular
configuration. Using time-delay interferometry (TDI), three noise-orthogonal
channels, denoted $A$, $E$, and $T$, can be constructed for each detector.
Among them, the $A$ and $E$ channels are sensitive to gravitational waves in
the frequency range relevant for this work, while the $T$ channel becomes
insensitive in the low-frequency limit~\cite{Orlando:2020oko}. We therefore
construct the detector network using the four channels $A_L$, $E_L$, $A_T$,
$E_T$, where the subscripts ``$L$'' and ``$T$'' denote LISA and TAIJI,
respectively.

The observed data stream in channel $i$ is modeled in the frequency domain as
\begin{equation}
  \tilde d_i(f) = \tilde s_i(f) + \tilde n_i(f),
  \label{eq:data-stream}
\end{equation}
where $\tilde s_i(f)$ and $\tilde n_i(f)$ represent the gravitational-wave
signal and instrumental noise, respectively, and the subscript $i$ labels the
observation channels $A_L$, $E_L$, $A_T$, $E_T$. The detector response to
gravitational waves is written as
\begin{equation}
  \tilde s_i(f) = D_i^{ab}(f)\, \tilde h_{ab}(f),
  \label{eq:detector-response}
\end{equation}
where $D_i^{ab}$ is the detector response tensor determined by the detector
geometry (see Appendix~\ref{app:detector-response} for explicit expressions).
Assuming stationary Gaussian instrumental noise, the noise correlation
satisfies
\begin{equation}
  \langle \tilde n_i(f)\, \tilde n_j^*(f') \rangle
  = \frac{1}{2}\, \delta_{ij}\, \delta(f-f')\, N_i(f),
  \label{eq:noise-correlation}
\end{equation}
where $N_i(f)$ is the one-sided noise power spectral density (PSD) of
channel $i$.

The SGWB signal appears coherently in the different detectors, while the
instrumental noises are uncorrelated between different detectors and
channels, so that $\langle \tilde n_i(f)\tilde n_j^*(f)\rangle = 0$ for
$i\ne j$. We therefore focus on the cross-correlation between distinct
channel pairs. The expectation value of the correlation estimator, averaged
over the observational time $T_{\rm obs}$, is
\begin{equation}
  \langle C_{ij}(f) \rangle \equiv \frac{1}{T_{\rm obs}} \left\langle \tilde d_i(f)\, \tilde d_j^*(f) \right\rangle
  = \frac{1}{T_{\rm obs}} \left\langle \tilde s_i(f)\, \tilde s_j^*(f) \right\rangle, \qquad i\ne j.
  \label{eq:cross-correlation-estimator}
\end{equation}
Following the overlap-reduction formalism for a polarized
SGWB~\cite{Chen:2024fto}, the expectation value of the cross-correlation can
be expressed in terms of the Stokes parameters $I(f)$ and $V(f)$ introduced
in Sec.~\ref{sec:polarized-sgwb} as
\begin{equation}
  \langle \tilde s_i(f)\, \tilde s_j^*(f) \rangle
  = \frac{1}{2}\left(\frac{3}{10}\right) T_{\rm obs} \left[ \gamma^I_{ij}(f)\, I(f) + \gamma^V_{ij}(f)\, V(f) \right],
  \label{eq:cross-correlation-stokes}
\end{equation}
where $\gamma^I_{ij}(f)$ and $\gamma^V_{ij}(f)$ are the overlap reduction
functions for the intensity and circular polarization, respectively, which
encode the relative detector geometry and separation. Their explicit forms
are given by
\begin{align}
  \gamma^I_{ij}(f) &= \frac{5}{2}\, D_i^{ab} D_j^{cd}
  \int \frac{d^2\hat{\vb*{k}}}{4\pi}
  \left[ e^+_{ab}(\hat{\vb*{k}})\, e^+_{cd}(\hat{\vb*{k}}) + e^\times_{ab}(\hat{\vb*{k}})\, e^\times_{cd}(\hat{\vb*{k}}) \right]
  e^{-i\hat{\vb*{k}}\cdot\Delta\vb*{r}/c}, \label{eq:gamma-I} \\
  \gamma^V_{ij}(f) &= \frac{5}{2}\, D_i^{ab} D_j^{cd}
  \int \frac{d^2\hat{\vb*{k}}}{4\pi} (-i)
  \left[ e^+_{ab}(\hat{\vb*{k}})\, e^\times_{cd}(\hat{\vb*{k}}) - e^\times_{ab}(\hat{\vb*{k}})\, e^+_{cd}(\hat{\vb*{k}}) \right]
  e^{-i\hat{\vb*{k}}\cdot\Delta\vb*{r}/c}, \label{eq:gamma-V}
\end{align}
where $\Delta\vb*{r}$ denotes the separation vector between the detector
pair. Detailed expressions for the LISA--TAIJI network, including the
detector tensors $D_i^{ab}$ and the noise PSDs $N_i(f)$, are summarized in
Appendix~\ref{app:detector-response}.

\subsection{Signal-to-noise ratio and PLI sensitivity}
\label{sec:snr-pli}

Assuming Gaussian instrumental noise, the likelihood for the measured
cross-correlation spectra is
\begin{equation}
  \mathcal{L}(C|\theta) \propto \exp\left[
  -\frac{1}{2} \sum_\kappa \int df\, \frac{|C_\kappa(f) - \langle C_\kappa(f;\theta)\rangle|^2}{\sigma_\kappa^2(f)}
  \right],
  \label{eq:likelihood}
\end{equation}
where $\theta$ denotes the set of parameters characterizing the SGWB
spectra, and $\kappa$ labels the independent detector-channel pairs used in
the cross-correlation. For the LISA--TAIJI network, the independent pairs are
$(A_L,A_T)$, $(A_L,E_T)$, $(E_L,A_T)$, and $(E_L,E_T)$. In the weak-signal
limit, the variance of the cross-correlation estimator is dominated by
instrumental noise,
\begin{equation}
  \sigma_\kappa^2(f) = \frac{N_i(f)\, N_j(f)}{2 T_{\rm obs}}.
  \label{eq:variance}
\end{equation}
Throughout this work, we assume that LISA and TAIJI have identical arm
lengths and comparable instrumental sensitivities, so that the effective
noise PSD for a detector pair can be written as
$N(f) = \sqrt{N_L(f)\, N_T(f)}$, where $N_L(f)$ and $N_T(f)$ denote the noise
PSDs of LISA and TAIJI, respectively (Appendix~\ref{app:detector-response}).

The SNRs for the SGWB intensity and circular-polarization components, for an
observation time $T_{\rm obs}$, are given by~\cite{Chen:2024fto}
\begin{align}
  {\rm SNR}_I^2 &= 2 T_{\rm obs} \left(\frac{3}{10}\right)^{\!2} \int_{f_{\min}}^{f_{\max}} df\,
  \left[\gamma^I_{\rm eff}(f)\right]^2 \frac{I^2(f)}{N^2(f)}, \label{eq:SNR-I} \\
  {\rm SNR}_V^2 &= 2 T_{\rm obs} \left(\frac{3}{10}\right)^{\!2} \int_{f_{\min}}^{f_{\max}} df\,
  \left[\gamma^V_{\rm eff}(f)\right]^2 \frac{V^2(f)}{N^2(f)}, \label{eq:SNR-V}
\end{align}
where $f_{\min}$ and $f_{\max}$ denote the minimum and maximum frequencies of
the observation band. Throughout this paper we adopt
\begin{equation}
  f_{\min} = 10^{-5}~{\rm Hz}, \qquad f_{\max} = 2\times10^{-2}~{\rm Hz},
  \label{eq:freq-range}
\end{equation}
corresponding to the frequency range relevant for the LISA--TAIJI network.
Here $\gamma^I_{\rm eff}(f)$ and $\gamma^V_{\rm eff}(f)$ are the effective
overlap reduction functions for the intensity and circular-polarization
modes, respectively, obtained by combining the contributions from all
independent channel pairs after orthogonalizing the mutual contamination
between the two modes,
\begin{align}
  \gamma^I_{\rm eff}(f) &= \sqrt{ \sum_\kappa \left[\gamma^I_\kappa(f)\right]^2
  - \frac{\left(\sum_\kappa \gamma^I_\kappa(f)\gamma^V_\kappa(f)\right)^2}{\sum_\kappa \left[\gamma^V_\kappa(f)\right]^2} }, \label{eq:gamma-I-eff} \\
  \gamma^V_{\rm eff}(f) &= \sqrt{ \sum_\kappa \left[\gamma^V_\kappa(f)\right]^2
  - \frac{\left(\sum_\kappa \gamma^I_\kappa(f)\gamma^V_\kappa(f)\right)^2}{\sum_\kappa \left[\gamma^I_\kappa(f)\right]^2} }. \label{eq:gamma-V-eff}
\end{align}

To visualize the sensitivity of the detector network to SGWBs with various
power-law spectra, we adopt the power-law integrated (PLI) sensitivity
curve~\cite{Thrane:2013oya}. For a power-law SGWB spectrum,
\begin{equation}
  \Omega_\mathrm{gw}(f) = \Omega^{I,V}_\alpha \left(\frac{f}{f_{\rm ref}}\right)^{\!\alpha},
  \label{eq:power-law-spectrum}
\end{equation}
the amplitude corresponding to a threshold ${\rm SNR}_{\rm th}$ is
\begin{equation}
  \Omega^{I,V}_\alpha = \frac{10}{3}\, \frac{{\rm SNR}_{\rm th}}{\sqrt{T_{\rm obs}}}\,
  \frac{4\pi^2}{3H_0^2}
  \left[ 2 \int_{f_{\min}}^{f_{\max}} df\, \frac{\left[\gamma^{I,V}_{\rm eff}(f)\right]^2 f^{2\alpha-6}}{N^2(f)} \right]^{-1/2}
  f_{\rm ref}^\alpha,
  \label{eq:Omega-alpha}
\end{equation}
where $f_{\rm ref}=10^{-3}$~Hz is the reference frequency. The PLI
sensitivity curve is then obtained as the envelope of these detectable
power-law spectra over a range of spectral indices $\alpha$,
\begin{equation}
  \Omega^{I,V}_{\rm PLI}(f) = \max_\alpha \left[ \Omega^{I,V}_\alpha \left(\frac{f}{f_{\rm ref}}\right)^{\!\alpha} \right].
  \label{eq:PLI-envelope}
\end{equation}
In this work we take $\alpha\in[-4,4]$ and adopt $T_{\rm obs}=3~\mathrm{yr}$ for the PLI sensitivity curves shown in Sec.~\ref{sec:sensitivity-results}.

\subsection{Fisher forecast}
\label{sec:fisher}

We finally introduce the Fisher-matrix formalism used to estimate the
expected parameter constraints from the LISA--TAIJI network. For
Gaussian-distributed data, the Fisher information matrix is defined by the
ensemble average of the second derivative of the log-likelihood
\eqref{eq:likelihood},
\begin{equation}
  F_{ab} \equiv - \left\langle \frac{\partial^2 \ln\mathcal{L}}{\partial\theta_a \partial\theta_b} \right\rangle
  = 2 T_{\rm obs} \sum_\kappa \int_{f_{\min}}^{f_{\max}} df\,
  \frac{\partial_a \langle C_\kappa(f)\rangle\, \partial_b \langle C_\kappa(f)\rangle}{N_\kappa^2(f)},
  \label{eq:fisher-matrix}
\end{equation}
where $\partial_a \equiv \partial/\partial\theta_a$. Using
Eq.~\eqref{eq:cross-correlation-stokes}, the expectation value of the
cross-correlation signal is
\begin{equation}
  \langle C_\kappa(f) \rangle = \frac{1}{2}\left(\frac{3}{10}\right)
  \left[ \gamma^I_\kappa(f)\, I(f) + \gamma^V_\kappa(f)\, V(f) \right],
  \label{eq:C-kappa-expectation}
\end{equation}
so that Eq.~\eqref{eq:fisher-matrix} becomes
\begin{equation}
  F_{ab} = \frac{1}{2}\left(\frac{3}{10}\right)^{\!2} T_{\rm obs} \sum_\kappa \int_{f_{\min}}^{f_{\max}} df\,
  \frac{\left[\gamma^I_\kappa(f)\, \partial_a I(f) + \gamma^V_\kappa(f)\, \partial_a V(f)\right]
  \left[\gamma^I_\kappa(f)\, \partial_b I(f) + \gamma^V_\kappa(f)\, \partial_b V(f)\right]}{N_\kappa^2(f)}.
  \label{eq:fisher-matrix-explicit}
\end{equation}
The Fisher matrix corresponds to the inverse covariance matrix in the
Gaussian approximation, so that the marginalized $1\sigma$ uncertainty of a
parameter $\theta_a$ is estimated as $\sigma(\theta_a) = \sqrt{(F^{-1})_{aa}}$;
the off-diagonal components of $F^{-1}$ quantify the parameter degeneracies
between different SGWB parameters. In Sec.~\ref{sec:fisher-results} we apply
this Fisher analysis to the phenomenological PMF models of
Sec.~\ref{sec:pheno-models} and investigate how accurately the LISA--TAIJI
network can constrain the PMF amplitude $\mathcal{B}$ and the helicity
fraction $r_H$ through joint measurements of the SGWB intensity and circular
polarization.

\section{Detectability of helical PMFs with the LISA--TAIJI network}
\label{sec:detectability}

We now apply the formalism of Sec.~\ref{sec:lisa-taiji} to the phenomenological
PMF models of Sec.~\ref{sec:pheno-models} and assess how well the LISA--TAIJI
network can probe helical PMFs. We first present the planned sensitivity of
the network in Sec.~\ref{sec:sensitivity-results}), then perform a Fisher
forecast to evaluate its potential to identify the helical component of PMFs
in Sec.~\ref{sec:fisher-results}, and finally compute the SNR for the
polarization signal to derive a phenomenological detection criterion
in Sec.~\ref{sec:snr-results}.

For the LISA--TAIJI network, we consider two configurations for TAIJI,
denoted TAIJIp and TAIJIm, following the constellation geometries summarized
in Appendix~\ref{app:detector-response}. Sensitivities are derived from the
cross-correlation between the two detectors for each configuration
(LISA-TAIJIp, LISA-TAIJIm), and, for comparison, from the autocorrelation of
the LISA dataset alone (``LISA (auto)''); the latter is valid only when the
noise and the signal can be well modeled and separated.

\subsection{Sensitivity of the network}
\label{sec:sensitivity-results}

To illustrate the sensitivity of a given detector configuration, we adopt
the PLI sensitivity curve introduced in Sec.~\ref{sec:snr-pli}, with
observation time $T_{\rm obs}=3$~yr and threshold ${\rm SNR}_{\rm th}=2$,
corresponding to a $2\sigma$ detection significance.

\begin{figure}
  \centering
\includegraphics[width=.47\textwidth]{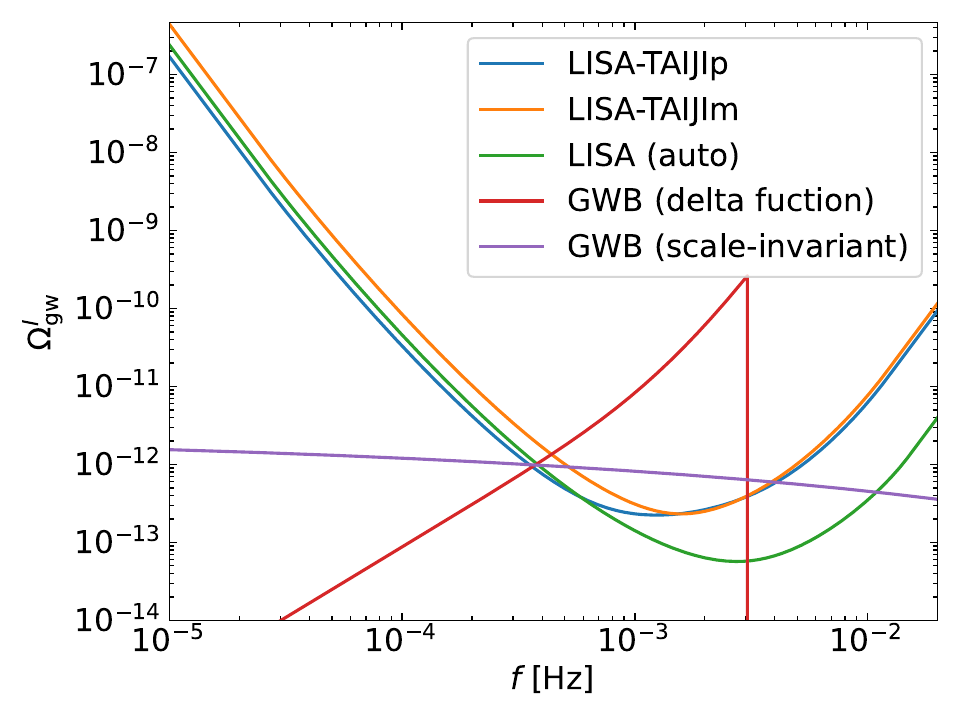}
\qquad
\includegraphics[width=.47\textwidth]{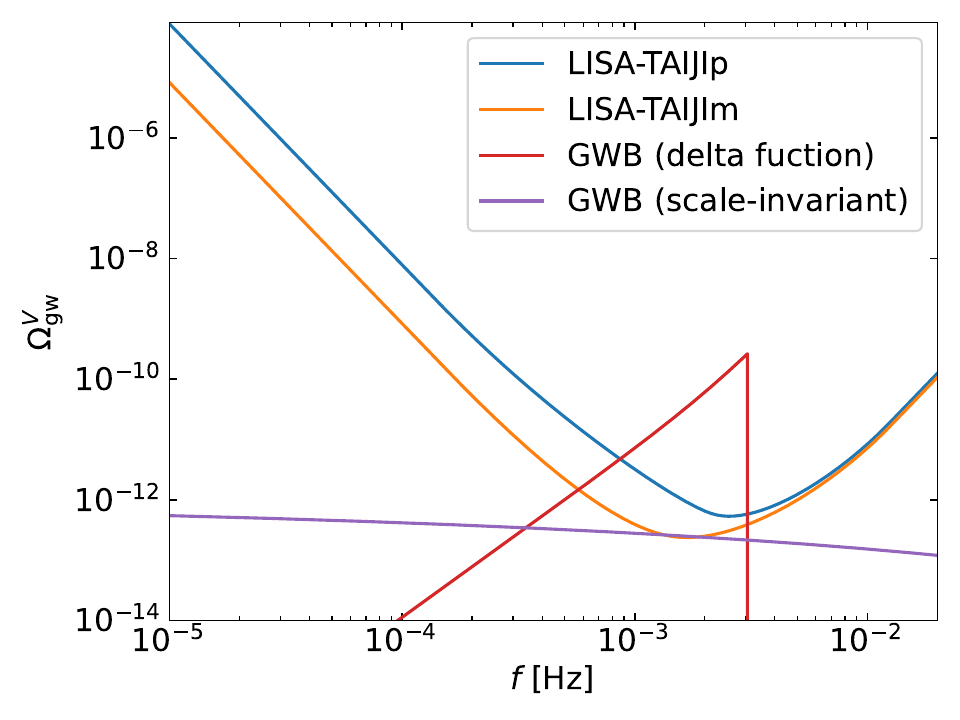}
  \caption{The power-law integrated (PLI) sensitivity curves for each
  detector configuration, assuming $T_{\rm obs}=3$~yr and
  ${\rm SNR}_{\rm th}=2$. Left: PLI sensitivities of LISA-TAIJIp,
  LISA-TAIJIm, and LISA (auto) for the intensity. Right: PLI sensitivities
  of LISA-TAIJIp and LISA-TAIJIm for the circular polarization; the PLI
  sensitivity for LISA alone does not appear, since a single planar
  detector has no sensitivity to circular polarization. For comparison, the
  GW spectra of the delta-function-type and scale-invariant-type PMFs with
  $(\mathcal{B},r_H)=(50~{\rm nG},1)$ are also shown.}
  \label{fig:pli-sensitivity}
\end{figure}

Figure~\ref{fig:pli-sensitivity} shows the resulting PLI sensitivities of the
LISA--TAIJI network and of LISA alone, for the intensity (left panel) and the
circular polarization (right panel). Since a single planar detector has no
sensitivity to circular polarization ($\gamma^V_{ii}=0$ for a vanishing
detector separation), the PLI sensitivity for LISA alone does not appear in
the right panel. For comparison, we overplot the GW spectra of the
delta-function-type and scale-invariant-type PMFs with
$(\mathcal{B},r_H)=(50~{\rm nG},1)$.

Regarding the intensity, LISA-TAIJIp is more sensitive than LISA-TAIJIm by a
factor in the low-frequency range, while the autocorrelation of LISA alone
achieves a much higher sensitivity than either network configuration. This
is because the overlap reduction function for the intensity,
$\gamma^I_{ij}(f)$, decreases as the physical separation between the two
detectors increases, and is maximal for the (unphysical) case of coincident
detectors realized by the LISA autocorrelation. Regarding the circular
polarization, LISA-TAIJIm achieves a higher sensitivity than LISA-TAIJIp by
an order of magnitude; conversely, the overlap reduction function for the
circular polarization, $\gamma^V_{ij}(f)$, vanishes identically when the two
detectors are colocated, so the LISA autocorrelation carries no sensitivity
to $V(f)$ at all. Hence, although the LISA--TAIJI network is less sensitive
to the intensity than the LISA autocorrelation, it offers unique information
on the circular polarization that is inaccessible to any single planar
detector.

\subsection{Fisher forecast for $\mathcal{B}$ and $r_H$}
\label{sec:fisher-results}

Based on the sensitivities illustrated above, we discuss the extent to which
the LISA--TAIJI network can probe helical PMFs. In the following, we fix the
conformal time at PMF generation to $\eta_B=10^{-12}\eta_\nu$, the peak scale
to $k_\mathrm{p}=10^{12}~{\rm Mpc}^{-1}$ for the delta-function-type PMFs; the energy spectrum of the SGWB is then
completely characterized by the two parameters $\mathcal{B}$ and $r_H$. We
adopt LISA-TAIJIm as the fiducial network configuration, since it is the
most sensitive to the circular polarization (Sec.~\ref{sec:sensitivity-results}),
and take the observation time to be $T_{\rm obs}=3$~yr.

\begin{figure}
  \centering
\includegraphics[width=.47\textwidth]{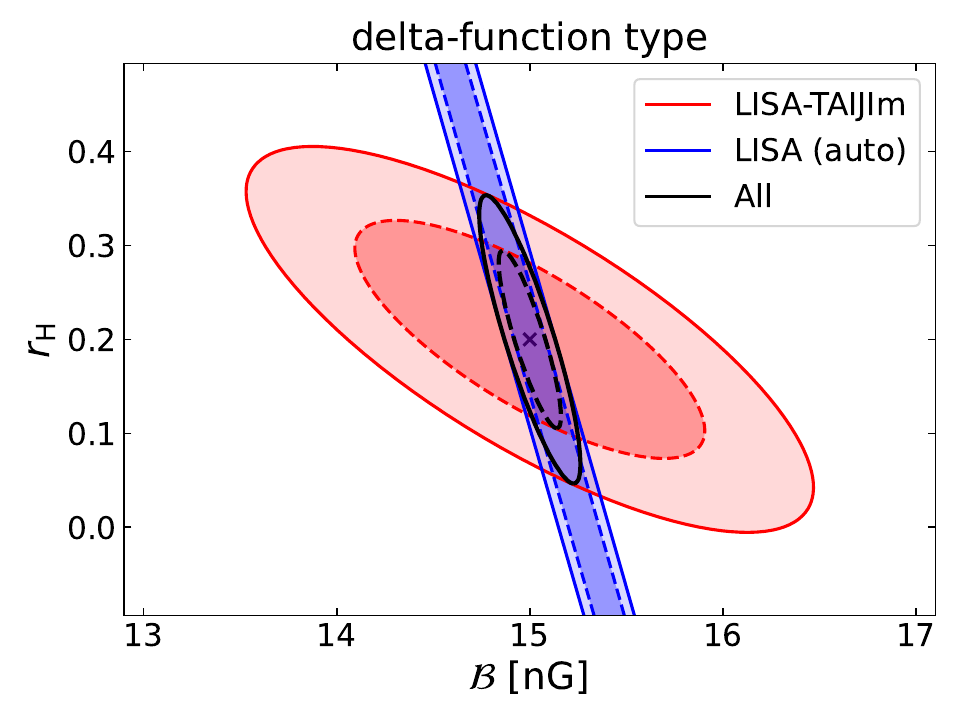}
\qquad
\includegraphics[width=.47\textwidth]{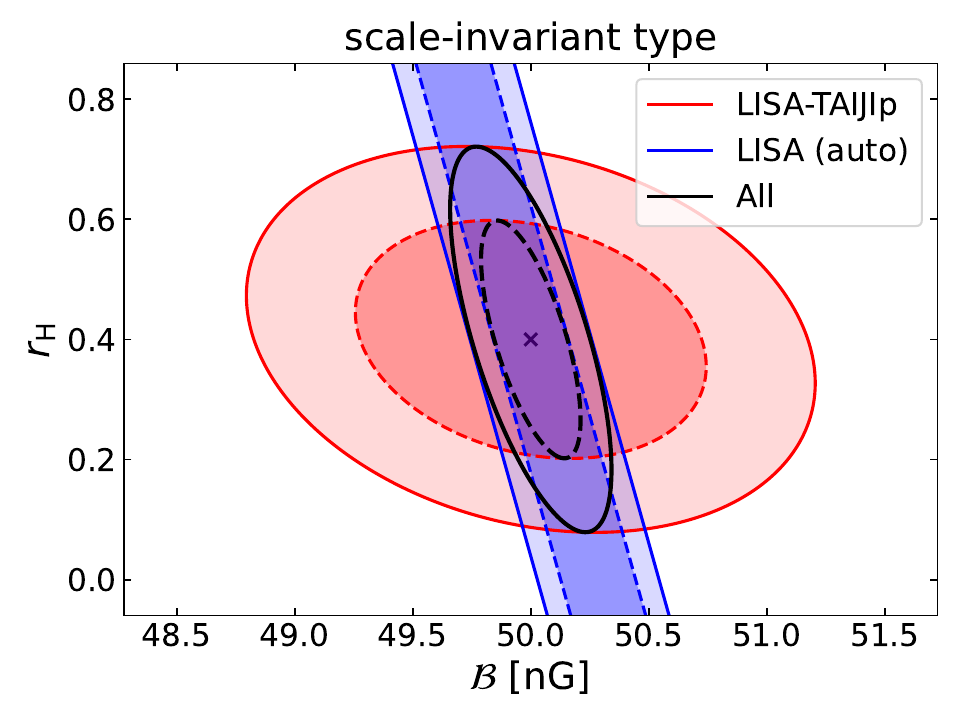}
  \caption{Results of the Fisher forecast for the two PMF models. The panels
  show the $1\sigma$ and $2\sigma$ uncertainties of $\mathcal{B}$ and $r_H$
  for the delta-function-type (left) and scale-invariant-type (right)
  models. Red and blue contours show the uncertainties obtained from
  LISA-TAIJIm and from LISA (auto), respectively; the black contour shows
  the combined constraint. The injected parameters (black cross) are
  $(\mathcal{B},r_H)=(15~{\rm nG},0.2)$ and $(50~{\rm nG},0.4)$,
  respectively.}
  \label{fig:fisher-forecast}
\end{figure}

Figure~\ref{fig:fisher-forecast} shows the results of the Fisher forecast of
Sec.~\ref{sec:fisher}, for the two phenomenological PMF models, comparing the
parameter estimation errors obtained from LISA alone and from the
LISA--TAIJI network. In the left panel, we inject the delta-function-type
PMFs with $(\mathcal{B},r_H)=(15~{\rm nG},0.2)$. The blue contour shows a
strong degeneracy between $\mathcal{B}$ and $r_H$ in the estimation from
LISA alone: since LISA is sensitive only to the GW intensity, and
$f(k)\propto\mathcal{B}^4[\,\cdots + 4r_H^2(\cdots)^2\,]$
(from Eq.~\eqref{eq:fg-delta-fS} and~\eqref{eq:fg-delta-fA}) depends on $\mathcal{B}$ and $r_H$ only through a
combination that is nearly degenerate for $|r_H|\lesssim1$, the intensity
alone cannot disentangle the two parameters. The red contour shows the
uncertainty from the LISA--TAIJI network, which additionally measures the
circular polarization; since $g(k)\propto r_H\mathcal{B}^4$
(from Eq.~\eqref{eq:fg-delta-g}) has a different --- and stronger --- dependence
on $r_H$ than $f(k)$ does, the direction of the resulting degeneracy differs
from that of the intensity-only measurement, enabling $\mathcal{B}$ and
$r_H$ to be jointly identified. In the right panel, we inject the
scale-invariant-type PMFs with $(\mathcal{B},r_H)=(50~{\rm nG},0.4)$. Since
the helical component does not significantly affect the intensity for this
model either, the directions of the degeneracies are slightly different from
the delta-function-type case, but the same conclusion holds. In both cases,
combining the LISA autocorrelation with the LISA--TAIJI cross-correlation
yields the tightest joint constraint on $\mathcal{B}$ and $r_H$ (black
contour). These results demonstrate that the LISA--TAIJI network can break
the degeneracy between the helical and non-helical contributions to the PMF
spectrum that a single planar detector cannot resolve.

\subsection{SNR criterion for helicity detection}
\label{sec:snr-results}

\begin{figure}
  \centering
\includegraphics[width=.47\textwidth]
{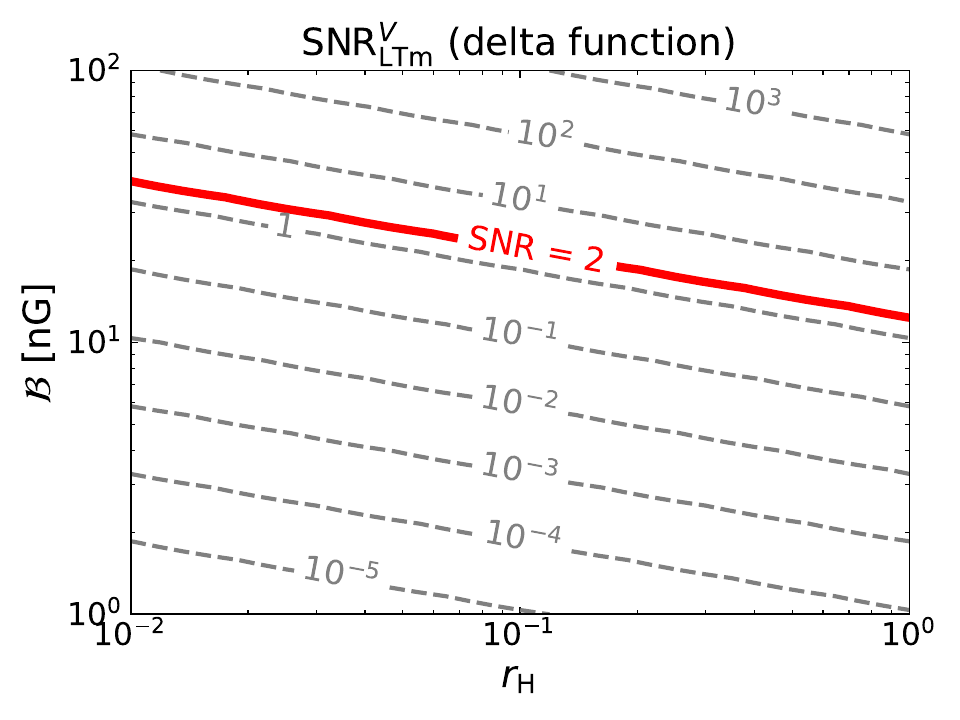}
\qquad
\includegraphics[width=.47\textwidth]{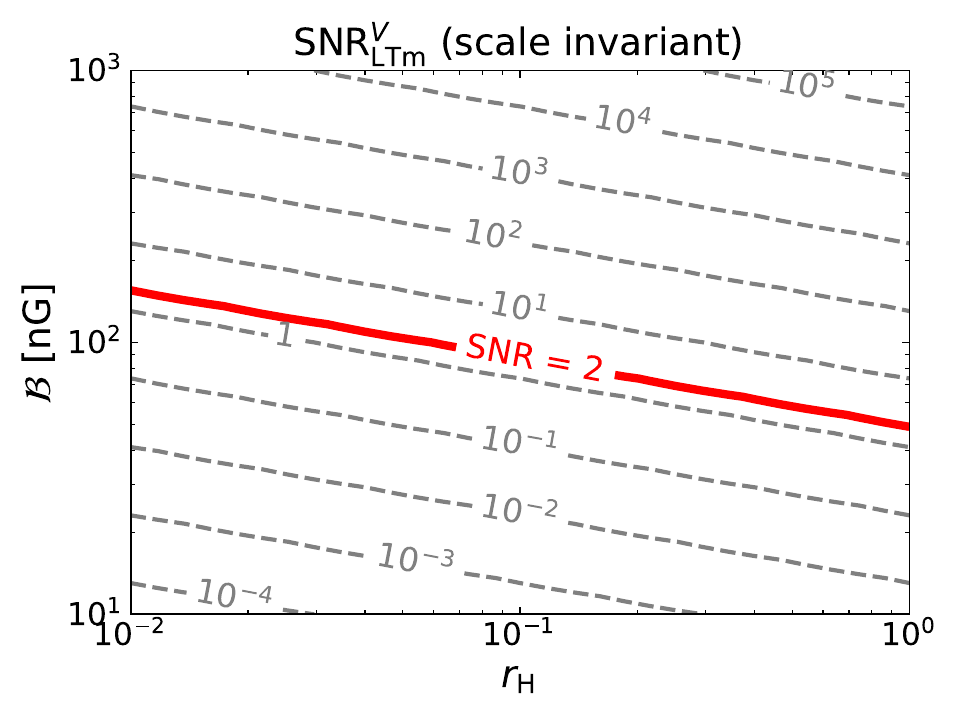}
  \caption{Signal-to-noise ratio for the GW circular polarization,
  ${\rm SNR}_V$, for the two PMF models observed with LISA-TAIJIm. Dashed
  contours show the level of ${\rm SNR}_V$ for various $(\mathcal{B},r_H)$
  for the delta-function-type (left) and scale-invariant-type (right)
  models. The solid red line indicates ${\rm SNR}_V=2$, corresponding to a
  $2\sigma$ detection.}
  \label{fig:snr-contours}
\end{figure}

Finally, we compute the SNR for each PMF model and derive a phenomenological
criterion for detecting the imprint of helical PMFs. Since the GW intensity
is mainly determined by the PMF amplitude $\mathcal{B}$ alone
(Sec.~\ref{sec:pheno-spectra}), we focus on ${\rm SNR}_V$, evaluated with the
fiducial LISA-TAIJIm configuration using Eq.~\eqref{eq:SNR-V}. Figure
\ref{fig:snr-contours} shows contours of ${\rm SNR}_V$ in the
$(\mathcal{B},r_H)$ plane for the two PMF models. Our formulation allows the
scaling of the SNR with $\mathcal{B}$ and $r_H$ to be derived analytically;
we find the fitting formulas
\begin{align}
  {\rm SNR}^V_{\rm LTm} &\simeq 2.4 \left(\frac{\mathcal{B}}{13~{\rm nG}}\right)^{\!4} r_H
  \qquad \text{(delta-function-type)}, \label{eq:SNR-fit-delta}\\
  {\rm SNR}^V_{\rm LTm} &\simeq 2.1 \left(\frac{\mathcal{B}}{50~{\rm nG}}\right)^{\!4} r_H
  \qquad \text{(scale-invariant-type)}. \label{eq:SNR-fit-SI}
\end{align}
The quartic scaling with $\mathcal{B}$ follows directly from
$V(f)\propto g(k)\propto r_H\mathcal{B}^4$
(Eqs.~\eqref{eq:V-stokes},~\eqref{eq:fg-delta-g}) and ${\rm SNR}_V\propto
|V(f)|$ at fixed spectral shape (Eq.~\eqref{eq:SNR-V}), while the linear
scaling with $r_H$ reflects the same proportionality directly.

As a reference, we plot the level corresponding to a $2\sigma$ detection
(${\rm SNR}_V=2$) as the red lines in Fig.~\ref{fig:snr-contours}. These
lines indicate that detecting the circular polarization requires at least
$\mathcal{B}\simeq10$~nG (delta-function type) or $\mathcal{B}\simeq50$~nG
(scale-invariant type) for fully helical PMFs, $r_H=1$; smaller values of
$r_H$ require correspondingly larger $\mathcal{B}$. Some theoretical models
of parity-violating inflationary magnetogenesis predict fully helical PMFs
as their natural outcome~\cite{Caprini:2014mja,Fujita:2019pmi,Sharma:2018kgs},
so this criterion provides a meaningful observational benchmark for such
scenarios. The required PMF strength is about one order of magnitude larger
than the current upper limit derived from Planck 2015~\cite{Planck:2015zrl};
however, the LISA--TAIJI network probes PMFs on comoving scales
$k_{\rm LISA}\approx10^{12}~{\rm Mpc}^{-1}$ that are entirely inaccessible to
CMB observations, and therefore offers unique and complementary information
about the PMFs.

\section{Summary}\label{sec:summary}
PMF is a promising candidate for generating a stochastic GWB. Parity violation in PMF production induces a helical component, imprinting its signature on the circular polarization of GWB. The LISA-TAIJI network is proposed to observe the circular polarization of GWs and is expected to offer valuable insights into new physics related to parity violation.
In advance of this future observation, in this study, we investigated the possibility of probing the imprint of helical PMFs by adopting the planned sensitivity. 
We considered the presence of helical PMFs produced during inflationary magnetogenesis and introduced several types of PMF power spectra phenomenologically.
Based on this formulation, we performed a Fisher forecast and calculated the SNR to estimate parameter constraints on helical PMFs.

As a result, we confirm that the resultant intensity does not depend significantly on the helicity fraction $r_H$. In contrast, circular polarization depends linearly on $r_H$, reflecting the existence of a helical component of PMFs.
Furthermore, our Fisher forecast results suggest that the LISA-TAIJI network can potentially break the degeneracy between the helical and non-helical components of PMFs, while a single planar detector cannot resolve it.
Lastly, we have numerically estimated the SNR for circular polarization and derived the lower limit on the magnitude of the comoving magnetic fields $\mathcal{B}$: $\mathcal{B}\gtrsim 10~\mathrm{nG}$ for delta-function-type PMFs and $\mathcal{B}\gtrsim 50~\mathrm{nG}$ for scale-invariant-type PMFs.
This is the unique prediction of the helical PMFs on the small scale $k=k_\mathrm{LISA}\approx10^{12}~\mathrm{Mpc}^{-1}$ and offers a limit to validate the theoretical models that predict the existence of helical PMFs as the parity-violating signature.

As future work, we can also constrain helical PMFs using different GW observations, such as pulsar timing arrays (PTAs). However, PTA observations are insensitive to the circular polarization of the isotropic GWB~\cite{Kato:2015bye}. Therefore, we need to focus on the anisotropic component of the GWB~\cite{Sato-Polito:2021efu,Cruz:2024svc}, consider polarimetry observations as proposed in Ref.~\cite{Liang:2025vji}, or a dipole PTA measurement recently suggested in Ref~\cite{Xu:2026ltq}.
Chiral GWs generated by helical PMFs can also affect polarization in CMB measurements.
In particular, the EB modes are sensitive to the circular polarization of GWs~\cite{Fujita:2022qlk} and thus to helical PMFs~\cite{Caprini:2003vc}; however, this has not yet been tested using Planck 2015 data.
Furthermore, the galaxy four-point cross-correlation can also capture parity violation, and recent BOSS observations suggest evidence for parity-violating signatures~\cite{Philcox:2021hbm,Philcox:2022hkh}. Such observations open up the possibility of constraining helical PMFs at galactic scales~\cite{Yura:2025mus}.

By combining observational constraints across multiple scales and theoretical knowledge from high-energy physics (see e.g., Ref.~\cite{Kamada:2020bmb}), we can achieve a comprehensive understanding of the origin of cosmic magnetic fields and parity-violating signatures in the early universe.

\acknowledgments
We thank Shohei Saga and Kohei Kamada for useful discussions. KF also thanks Ju Chen for assistance in reproducing the formulation for the LISA-TAIJI network.
KF is supported by Japan Society for the Promotion of Science (JSPS) KAKENHI Grant Number JP25KJ1388.

\appendix

\section{Derivation of $f(k)$ and $g(k)$ for the phenomenological PMF models}
\label{app:fg-derivation}

In this appendix, we derive the parity-even and parity-odd anisotropic-stress
spectra $f(k)$ and $g(k)$, defined by the general convolution integrals
\eqref{eq:f-general}--\eqref{eq:g-general}, for the two phenomenological PMF
models introduced in Sec.~\ref{sec:pheno-models}: the delta-function-type
spectrum~\eqref{eq:delta-type} (Appendix~\ref{app:delta-derivation}) and the scale-invariant spectrum~\eqref{eq:scale-inv-type}
(Appendix~\ref{app:scale-inv-derivation}).
For convenience, we split the intensity source $f(k)$ into $f(k)=f^S(k)+f^A(k)$ as
\begin{align}
      f^S(k) &=  \frac{3^2}{4(4\pi)^2\rho_{\gamma,0}^2} \int \frac{d^3p}{(2\pi)^3}\Big[ (1+\gamma^2)(1+\beta^2)\, S(p)\,S(|\vb*{k}-\vb*{p}|)\Big],\\
      f^A(k) &=  \frac{3^2}{4(4\pi)^2\rho_{\gamma,0}^2} \int \frac{d^3p}{(2\pi)^3}\Big[4\gamma\beta\, A(p)\,A(|\vb*{k}-\vb*{p}|)\Big].
\end{align}
which denote the contributions to the intensity from the non-helical and helical components of PMFs, respectively.

\subsection{delta-function-type PMFs}
\label{app:delta-derivation}

For the delta-function-type spectrum, $S(p)$ and $A(p)$ are both proportional to $\delta_\mathrm{D}(\ln(p/k_\mathrm{p}))$, i.e., they vanish unless $|\vb*{p}|=k_\mathrm{p}$. For instance, by substituting the delta-function-type spectrum into $f^S(k)$, the convolution integral is described as 
\begin{equation}
\begin{split}
    &f^S(k) \sim \int_0^\infty \dd p~p^{-1} \delta_\mathrm{D}[\ln(p/k_\mathrm{p})]\int_{-1}^{1}\dd \gamma~|\vb*{k-p}|^{-3}(1+\gamma^2)(1+\beta^2)\delta_\mathrm{D}\Big[\ln\Big(\frac{|\vb*{k-p}|}{k_\mathrm{p}}\Big)\Big]\\
    &=\int_0^\infty \dd p~\Big(\frac{k_\mathrm{p}}{p}\Big)\,\delta_\mathrm{D}(p-k_\mathrm{p})\int_{-1}^{1}\dd \gamma~|\vb*{k-p}|^{-3}(1+\gamma^2)(1+\beta^2)\,\Big(\frac{k_\mathrm{p}^2}{kp}\Big)\,\delta_\mathrm{D}\Big(\gamma-\frac{k^2+p^2-k_\mathrm{p}^2}{2kp}\Big)
\end{split}
\end{equation}
where, $\gamma=\hat{\vb*{k}}\cdot\hat{\vb*{p}}$,~$|\vb*{k-p}|=\sqrt{k^2-2kp\gamma+p^2}$ and $\beta = \hat{\vb*{k}}\cdot\widehat{\vb*{k}-\vb*{p}} = (k-p\gamma)/\sqrt{k^2-2kp\gamma+p^2}$.  
Note that in the second line, we apply the properties of the delta function, which are $\delta_\mathrm{D}(\alpha x)=\delta_\mathrm{D}(x)/|\alpha|$ and $\delta_\mathrm{D}(f(x))=\sum_i\delta_\mathrm{D}(x-a_i)/|f^\prime(a_i)|$ where $a_i$ is the $i$-th solution of a function $f(x)$. 
The convolution integral above requires \emph{both} $|\vb*{p}|=k_\mathrm{p}$ \emph{and} $|\vb*{k}-\vb*{p}|=k_\mathrm{p}$ simultaneously. The solution satisfying these conditions is therefore
\begin{equation}
    \gamma = \frac{k^2+p^2-k_\mathrm{p}^2}{2kp} = \frac{k}{2k_\mathrm{p}}
  \label{eq:law-of-cosines}
\end{equation}
and this solution holds only for $k\le 2k_\mathrm{p}$. In this case, $\beta = k/2k_\mathrm{p}$ because two vectors $\vb*{p}$ and $\vb*{k-p}$ form an isosceles triangle. For $k>2k_\mathrm{p}$, since $|\vb*{p}|=k_\mathrm{p}$ and $|\vb*{k}-\vb*{p}|=k_\mathrm{p}$ can not be satisfied simultaneously, the convolution integral is zero. By combining these conditions and all overall factors, we can derive the explicit form of $f^S(k)$ as 
\begin{equation}
\label{eq:fg-delta-fS}
    f^S(k) = \frac{9}{64}\Big(\frac{\mathcal{B}^2}{\rho_{\gamma,0}}\Big)^2k_\mathrm{p}^{-2}\frac{1}{k}\Big(1+\frac{k^2}{4k_\mathrm{p}^2}\Big)^2\Theta_\mathrm{H}\Big(1-\frac{k}{2k_\mathrm{p}}\Big).
\end{equation}
where, $\Theta_\mathrm{H}(x)$ indicates Heaviside step function. In the same way, $f^A(k)$ and $g(k)$ can be calculated by 
\begin{equation}
\label{eq:fg-delta-fA}
    f^A(k) = \frac{9}{64}\Big(\frac{\mathcal{B}^2}{\rho_{\gamma,0}}\Big)^2k_\mathrm{p}^{-2}\frac{4r_H^2}{k}\Big(\frac{k}{2k_\mathrm{p}}\Big)^2\Theta_\mathrm{H}\Big(1-\frac{k}{2k_\mathrm{p}}\Big),
\end{equation}
and 
\begin{equation}
\label{eq:fg-delta-g}
    g(k) = \frac{9}{64}r_H\Big(\frac{\mathcal{B}^2}{\rho_{\gamma,0}}\Big)^2k_\mathrm{p}^{-2}\frac{4}{k}\Big(\frac{k}{2k_\mathrm{p}}\Big)\Big(1+\frac{k^2}{4k_\mathrm{p}^2}\Big)\Theta_\mathrm{H}\Big(1-\frac{k}{2k_\mathrm{p}}\Big).
\end{equation}

\label{eq:fg-delta}
These expressions explicitly show that $f(k)\propto \mathcal{B}^4$, while
$g(k)\propto r_H\mathcal{B}^4$: the intensity source $f(k)=f^S(k)+f^A(k)$ receives contributions from both $S(p)S(|\vb*{k}-\vb*{p}|)$ and $A(p)A(|\vb*{k}-\vb*{p}|)$, whereas the polarization source $g(k)$ arises purely from the cross term
$S(p)A(|\vb*{k}-\vb*{p}|)$ and $A(p)S(|\vb*{k}-\vb*{p}|)$, and is therefore exactly linear in $r_H$. This is the analytic origin of the scaling quoted in Sec.~\ref{sec:pheno-spectra}.

\subsection{Scale-invariant PMFs}
\label{app:scale-inv-derivation}

For the scale-invariant spectrum~\eqref{eq:scale-inv-type}, the convolution
integrals \eqref{eq:f-general}--\eqref{eq:g-general} do not admit a closed
analytic form and must be evaluated numerically. Our computation follows the
method of Ref.~\cite{Caprini:2003vc}. To facilitate the calculation, we
define the auxiliary function
\begin{equation}
  F_{\alpha,\delta,l,m}(k) \equiv
  \int_{k_{\min}}^{k_{\max}} dp \int_{-1}^{1} d\gamma\,
  (k^2+p^2-2kp\gamma)^{\alpha/2}\, p^{\delta}\, k^{l}\, \gamma^{m},
  \label{eq:F-function}
\end{equation}
where $k$, $p$, and $\gamma=\hat{\vb*{k}}\cdot\hat{\vb*{p}}$ follow the same
definitions as in Sec.~\ref{sec:aniso-stress}. Integrating
$F_{\alpha,\delta,l,m}(k)$ over $\gamma$ analytically (by repeated
integration by parts) gives
\begin{equation}
\begin{split}
  F_{\alpha,\delta,l,m}(k) = \sum_{n=1}^{m+1}
  \left[ \frac{m!}{(m-n+1)!} \prod_{i=1}^{n} \frac{1}{\alpha+2i} \right]
  &\int_{k_{\min}}^{k_{\max}} dp\; k^{l} p^{\delta} (kp)^{-n}\\
  &\times\left[(-1)^{m-n+1}(k+p)^{\alpha+2n}- |k-p|^{\alpha+2n} \right],
\end{split}
\label{eq:F-integrated}
\end{equation}
where $n$ denotes the number of integrations by parts applied. To perform the
remaining integral over $p$, we approximate the integrand using a binomial
expansion in $k/p$ (for $k<p$) or $p/k$ (for $k>p$), keeping terms up to
cubic order. Writing $\alpha_*\equiv\alpha+2n$, when $m-n+1$ is even,
\begin{equation}
  (-1)^{m-n+1}(k+p)^{\alpha_*} - |k-p|^{\alpha_*} \approx
  \begin{cases}
    k^{\alpha_*}\left[ 2\alpha_* \dfrac{k}{p} + \dfrac{1}{3}\alpha_*(\alpha_*-1)(\alpha_*-2)\left(\dfrac{k}{p}\right)^{\!3} \right], & k<p, \\[2mm]
    p^{\alpha_*}\left[ 2\alpha_* \dfrac{p}{k} + \dfrac{1}{3}\alpha_*(\alpha_*-1)(\alpha_*-2)\left(\dfrac{p}{k}\right)^{\!3} \right], & k>p,
  \end{cases}
  \label{eq:binomial-even}
\end{equation}
while when $m-n+1$ is odd,
\begin{equation}
  (-1)^{m-n+1}(k+p)^{\alpha_*} - |k-p|^{\alpha_*} \approx
  \begin{cases}
    -k^{\alpha_*}\left[ 2 + \alpha_*(\alpha_*-1)\left(\dfrac{k}{p}\right)^{\!2} \right], & k<p, \\[2mm]
    -p^{\alpha_*}\left[ 2 + \alpha_*(\alpha_*-1)\left(\dfrac{p}{k}\right)^{\!2} \right], & k>p.
  \end{cases}
  \label{eq:binomial-odd}
\end{equation}

Using $F_{\alpha,\delta,l,m}(k)$ together with
$\beta = (k-p\gamma)/\sqrt{k^2+p^2-2kp\gamma}$, the parity-even source
$f(k)=f^S(k)+f^A(k)$ of Eq.~\eqref{eq:f-general} can be written as
\begin{align}
  f^S(k) &\sim \int_{k_{\min}}^{k_{\max}} dp \int_{-1}^{1} d\gamma\,
  (1+\gamma^2)(1+\beta^2) (k^2+p^2-2kp\gamma)^{n_{\rm SI}/2}\, p^{-n_{\rm SI}+2} \nonumber\\
  &= 2F_{n_{\rm SI},n_{\rm SI},0,0} + 2F_{n_{\rm SI},n_{\rm SI},0,2}
  - 2F_{n_{\rm SI}-2,n_{\rm SI}+2,0,0} + 2F_{n_{\rm SI}-2,n_{\rm SI}+2,0,4}, \label{eq:fS} \\
  f^A(k) &\sim \int_{k_{\min}}^{k_{\max}} dp \int_{-1}^{1} d\gamma\,
  (\gamma\beta) (k^2+p^2-2kp\gamma)^{n_{\rm SI}/2}\, p^{-n_{\rm SI}+2} \nonumber\\
  &= F_{n_{\rm SI}-1,n_{\rm SI},1,1} + F_{n_{\rm SI}-1,n_{\rm SI}+1,0,2}, \label{eq:fA}
\end{align}
and the parity-odd source is
\begin{align}
  g(k) &\sim \int_{k_{\min}}^{k_{\max}} dp \int_{-1}^{1} d\gamma\,
  \left[(1+\gamma^2)\beta + \gamma(1+\beta^2)\right]
  (k^2+p^2-2kp\gamma)^{n_{\rm SI}/2}\, p^{-n_{\rm SI}+2} \nonumber\\
  &= F_{n_{\rm SI}-1,n_{\rm SI},1,0} - F_{n_{\rm SI}-1,n_{\rm SI}+1,0,1}
  + F_{n_{\rm SI}-1,n_{\rm SI},1,2} + F_{n_{\rm SI}-1,n_{\rm SI}+1,0,3} \nonumber\\
  &\quad + 2F_{n_{\rm SI},n_{\rm SI},0,1} - F_{n_{\rm SI}-2,n_{\rm SI}+2,0,1} + F_{n_{\rm SI}-2,n_{\rm SI}+2,0,3},
  \label{eq:g-SI}
\end{align}
where $n_{\rm SI}$ is the spectral index of the scale-invariant PMFs. A
purely scale-invariant spectrum, $n_{\rm SI}=-3$ exactly, renders the
integrals \eqref{eq:fS}--\eqref{eq:g-SI} logarithmically divergent, as is
generically the case for scale-invariant PMF spectra~\cite{Fujita:2019pmi}.
We therefore regularize the numerical integration by taking
$n_{\rm SI}=-2.99$, slightly away from exact scale invariance, while keeping
the normalization of Eq.~\eqref{eq:scale-inv-type} fixed. 
In evaluating Eqs.~\eqref{eq:fS}--\eqref{eq:g-SI}, we retain only the leading-order terms in each component: for instance, in $f^S(k)$ we extract the terms proportional to $k^{2n_{\rm SI}+3}$ from $2F_{n_{\rm SI},n_{\rm SI},0,0}$ and $2F_{n_{\rm SI},n_{\rm SI},0,2}$. On the other hand, from $2F_{n_{\rm SI}-2,n_{\rm SI}+2,0,0}$ and $2F_{n_{\rm SI}-2,n_{\rm SI}+2,0,4}$, the terms proportional to $k^{2n_{\rm SI}+5}$ are obtained. However, because these terms cancel between themselves, $f^S(k)$ is accordingly described by $k^{2n_{\rm SI}+3}$.

\section{Supplements for the calculation of the detector's response}
\label{app:detector-response}

In this appendix we summarize the calculation of the detector tensor, the
overlap reduction functions $\gamma^I_{ij}(f)$ and $\gamma^V_{ij}(f)$
introduced in Sec.~\ref{sec:detector-response}, and the noise PSD of the
LISA--TAIJI network, following Ref.~\cite{Chen:2024ikn}.

\subsection{Detector geometry and detector tensors}

\begin{figure}
  \centering
   \includegraphics[width=\textwidth]{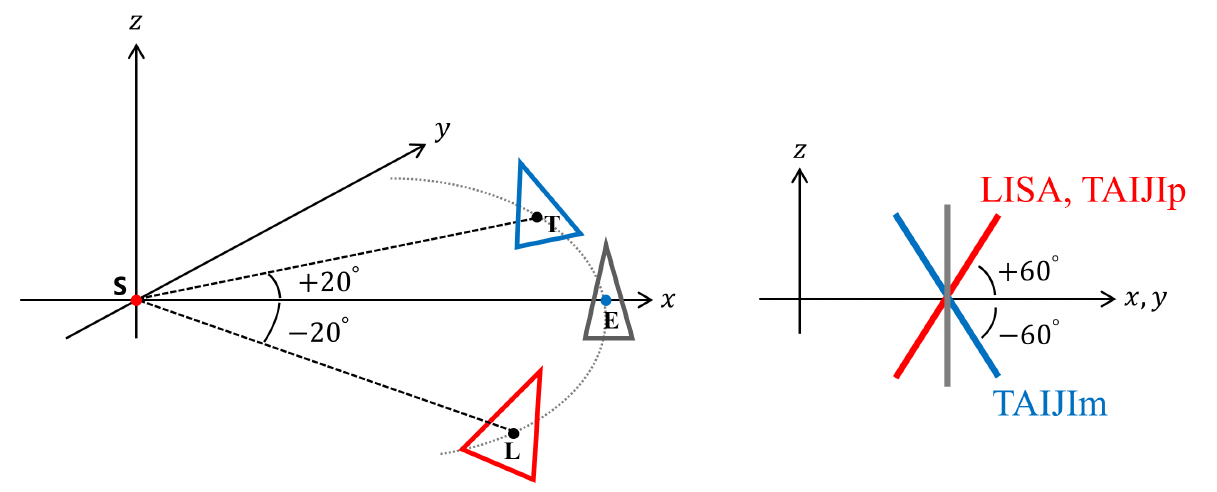}
  \caption{Schematic illustration of the relative positions and
  configurations of LISA and TAIJI. Left: the positions of the Sun (S),
  Earth (E), LISA (L), and TAIJI (T) in the ecliptic ($xy$) plane. Right:
  the inclinations of LISA and TAIJIp (red) and TAIJIm (blue). A
  hypothetical detector located at the Earth (grey) is shown in both panels
  for reference.}
  \label{fig:constellation}
\end{figure}

Figure~\ref{fig:constellation} shows the relative positions and
configurations of LISA and TAIJI. We take the $xy$-plane to be the ecliptic
plane, with the Sun at the coordinate origin and the Earth on the $x$-axis.
LISA and TAIJI travel along their heliocentric orbits together with the
Earth, keeping their angular separation from the Earth fixed at $-20^\circ$
and $+20^\circ$, respectively. To maintain a stable constellation, LISA and
TAIJIp are inclined by $60^\circ$, while TAIJIm is inclined by $-60^\circ$.

To compute a detector tensor, we consider a hypothetical detector located at
the Earth (the grey triangle in Fig.~\ref{fig:constellation}), initially
facing the Sun, so that the normal vector to its constellation plane is
$\vb*{n}_{\rm ini}=-\hat{\vb*{x}}$. Its detector tensors for the $A$ and $E$
channels are effectively described by two L-shaped interferometers differing
by a relative rotation of $45^\circ$,
\begin{equation}
  D_{\rm ini,A} = (\hat{\vb*{y}}\otimes\hat{\vb*{y}} - \hat{\vb*{z}}\otimes\hat{\vb*{z}})/2,
  \qquad
  D_{\rm ini,E} = (\hat{\vb*{y}}\otimes\hat{\vb*{z}} + \hat{\vb*{z}}\otimes\hat{\vb*{y}})/2.
  \label{eq:D-ini}
\end{equation}
The normal vectors of LISA (L), TAIJIp (Tp), and TAIJIm (Tm) are obtained by
rotating~$\vb*{n}_{\rm ini}$,
\begin{align}
  \vb*{n}_L &= R_L\, \vb*{n}_{\rm ini} = \left( -\tfrac{\sqrt3}{2}\cos20^\circ,\; \tfrac{\sqrt3}{2}\sin20^\circ,\; \tfrac12 \right), \nonumber\\
  \vb*{n}_{\rm Tp} &= R_{\rm Tp}\, \vb*{n}_{\rm ini} = \left( -\tfrac{\sqrt3}{2}\cos20^\circ,\; -\tfrac{\sqrt3}{2}\sin20^\circ,\; \tfrac12 \right), \label{eq:normal-vectors} \\
  \vb*{n}_{\rm Tm} &= R_{\rm Tm}\, \vb*{n}_{\rm ini} = \left( \tfrac{\sqrt3}{2}\cos20^\circ,\; \tfrac{\sqrt3}{2}\sin20^\circ,\; \tfrac12 \right), \nonumber
\end{align}
where $R_D$ is the rotation matrix realizing the configuration of detector
$D$: $R_L = R_z(-\pi/9)R_y(\pi/6)$, $R_{\rm Tp} = R_z(\pi/9)R_y(\pi/6)$, and
$R_{\rm Tm} = R_z(\pi/9)R_y(5\pi/6)$. Applying these rotations to the
initial detector tensors,
\begin{equation}
  D_i = R_D\, D_{{\rm ini},X}\, R_D^T,
  \label{eq:D-rotated}
\end{equation}
gives the detector tensor for channel $X\in\{A,E\}$ of detector
$D\in\{L,{\rm Tp},{\rm Tm}\}$, denoted $D_i$ using the channel index $i$.

\subsection{Overlap reduction functions}

Adopting these detector tensors, the overlap reduction functions for the
intensity ($I$) and circular polarization ($V$) between two channels $i,j$
can be computed as
\begin{equation}
  \gamma^{I,V}_{ij} = D_i^{ab} D_j^{cd}\, \hat\Gamma^{I,V}_{abcd}(\iota,\hat{\vb*{s}}),
  \label{eq:gamma-general}
\end{equation}
where $\hat{\vb*{s}}=\hat{\vb*{y}}$ is the unit vector along the separation of
the two detectors, and $\iota\equiv 2\pi\Delta r/c$. The parity-even kernel
is
\begin{equation}
  \hat\Gamma^I_{abcd}(\iota,\hat{\vb*{s}}) = b^I_0(\iota)\,\delta_{ab}\delta_{cd}
  + b^I_1(\iota)\,\delta_{ac}s_b s_d + b^I_2(\iota)\, s_a s_b s_c s_d,
  \label{eq:Gamma-I}
\end{equation}
with coefficients
\begin{equation}
  b^I_0(\iota) = 2j_0(\iota) - \frac{20}{7}j_2(\iota) + \frac{1}{7}j_4(\iota), \quad
  b^I_1(\iota) = \frac{60}{7}j_2(\iota) - \frac{10}{7}j_4(\iota), \quad
  b^I_2(\iota) = \frac{5}{2}j_4(\iota),
  \label{eq:b-I-coeffs}
\end{equation}
where $j_l(x)$ is the $l$-th spherical Bessel function of the first kind.
The parity-odd kernel is
\begin{equation}
  \hat\Gamma^V_{abcd}(\iota,\hat{\vb*{s}}) = b^V_0(\iota)\, w_{ac}\,\delta_{bd}
  + b^V_1(\iota)\, w_{ac}\, s_b s_d,
  \label{eq:Gamma-V}
\end{equation}
where $\equiv\epsilon_{abc}s^c$ and
\begin{equation}
  b^V_0(\iota) = 4j_1(\iota) - j_3(\iota), \qquad b^V_1(\iota) = 5j_3(\iota).
  \label{eq:b-V-coeffs}
\end{equation}

\subsection{Noise power spectral density}

For simplicity, we assume that each detector has equal arm lengths and that
each channel has the same noise properties. The noise PSD of detector $D$ is
then given by~\cite{Orlando:2020oko}
\begin{equation}
  N_D(f) = \frac{2}{3L_D^2} \left[
  1 + \frac{1}{2}\cos\!\left(\frac{f}{f_*}\right) \right] P_{\rm OMS}
  + 2\left[ 1 + \cos\!\left(\frac{f}{f_*}\right) + \cos^2\!\left(\frac{f}{f_*}\right) \right]
  \frac{P_{\rm acc}}{(2\pi f)^4},
  \label{eq:noise-PSD}
\end{equation}
where $L_D$ is the arm length of detector $D$ and $f_*\equiv c/(2\pi L_D)$.
In this work we adopt $L_L=L_T=2.5\times10^9$~m for LISA and TAIJI,
respectively. Here $P_{\rm OMS}$ is the optical metrology system noise and
$P_{\rm acc}$ is the test-mass acceleration noise,
\begin{align}
  P_{\rm OMS} &= A_{\rm OMS}^2 \left[ 1 + \left(\frac{2~{\rm mHz}}{f}\right)^{\!4} \right], \label{eq:P-OMS} \\
  P_{\rm acc} &= A_{\rm acc}^2 \left[ 1 + \left(\frac{0.4~{\rm mHz}}{f}\right)^{\!2} \right]
  \left[ 1 + \left(\frac{f}{8~{\rm mHz}}\right)^{\!2} \right], \label{eq:P-acc}
\end{align}
where $A_{\rm OMS}$ and $A_{\rm acc}$ are the amplitudes of each noise
source. For LISA, we adopt $A_{\rm OMS}=15~{\rm pm}/\sqrt{\rm Hz}$ and
$A_{\rm acc}=3~{\rm fm/s}^2/\sqrt{\rm Hz}$; for TAIJI, we adopt
$A_{\rm OMS}=8~{\rm pm}/\sqrt{\rm Hz}$ and
$A_{\rm acc}=3~{\rm fm/s}^2/\sqrt{\rm Hz}$.

\bibliographystyle{JHEP}
\bibliography{biblio.bib}

\end{document}